\documentclass{aastex62}
\shorttitle{The nucleosynthesis in short GRB outflows}
\shortauthors{Janiuk}

\begin{document}
\title{The r-process nucleosynthesis in the 
  outflows from short GRB accretion disks}

\correspondingauthor{A. Janiuk}
\email{agnieszka.janiuk@gmail.com}



\newcommand{\lrz}{\gamma}
\newcommand{\enth}{\xi}
\newcommand{\hslope}{h}
\newcommand{\polind}{\hat \Gamma}



\author{Agnieszka Janiuk}
\affil{Center for Theoretical Physics, Polish Academy of Sciences, 
Al. Lotnikow 32/46, 
02-668 Warsaw, Poland}
\email{agnes@cft.edu.pl}

\begin{abstract}
Short gamma-ray bursts require a rotating black hole, surrounded by a magnetized 
relativistic accretion disk, such as the one formed by coalescing binary
neutron stars or neutron star - black hole systems.
The accretion onto a Kerr black hole is the mechanism of launching a baryon-free
relativistic jet. 
An additional uncollimated outflow, consisting of sub-relativistic neutron-rich 
material which becomes unbound by thermal, magnetic and viscous forces, is responsible 
for blue and red kilonova.
We explore the formation, composition and geometry of the secondary outflow
 by means of simulating accretion disks with relativistic magneto-hydrodynamics
and employing realistic nuclear equation of state.
We calculate the nucleosynthetic r-process yields by sampling the outflow with a 
dense set of tracer particles.
Nuclear heating from the residual r-process radioactivities in the freshly synthesized nuclei 
is expected to power a red kilonova,
contributing independently from the 
dynamical ejecta component, launched at the time of merger, and neutron-poor broad polar 
outflow, launched from the surface of the hypermassive neutron star by neutrino wind.
 Our simulations show that both magnetisation of the disk and high black hole spin are able to launch fast wind outflows ($v/c\sim 0.11-0.23$) with a broad range of electron fraction $Y_{\rm e}\sim 0.1-0.4$, and help explain the multiple components observed in the kilonova lightcurves. The total mass loss from the post-merger disk via unbound outflows is between 2\% and 17\% of the initial disk mass.
\end{abstract}




\section{Introduction}

The compact object binaries, namely the Black Hole-Neutron Star (BHNS) and the Neutron Star-Neutron Star (NSNS) binaries \citep{Eichler_Livio_Piran_schramm_1989_Nature, Paczynski_1991_Acta_Astronomica, Narayanetal1992}, are favorable
progenitors for the Short Gamma Ray Bursts (sGRB) (see however \citet{2011NewAR..55....1B} and references therein for alternative explanations, including magnetars formed through a massive star core-collapse, or binary white dwarf mergers, or the white dwarf accretion-induced collapse, or accretion-induced collapse of neutron stars).
The complex nature of macroscopic and microphysical properties of the central engines requires a still ongoing effort to identify crucial aspects of their operation.
Some of the fundamental requirements for the mechanism responsible for the outflow launching were already stated years ago.
Apart from the collimated jets \citep{Sarietal1999, Rhoads1999}, these engines
are also supposed to produce the uncollimated, equatorial wind outflows,
mediated by magnetic fields and centrifugal forces \citep{2012MNRAS.426.3241N, Fernandez2015}.

A major breakthrough occurred recently with the multi-messenger
detection of the GRB 170817A through the VIRGO/LIGO and Fermi-GBM observatories. The waveform of the emitted gravitational waves is consistent with the merging of an NSNS
system \citep{Abbott2017}.
The event was followed by a sGRB after $\sim 1.7 s$, as given by the
Fermi Gamma-ray Space Telescope trigger.
The associated GRB was a few orders of magnitude fainter than a typical short burst, and its energy $E_{\rm iso}=\left(3.1\pm0.7\right)\cdot 10^{46}$ ergs was explained as the off jet-axis observation and the line of sight passing
through the surrounding cocoon \citep{Lazzati2017}, or an intrinsic property of the specific NSNS merging event \citep{Murguia_Berthier2017, Zhang_et_al_2017arXiV}.
The optical counterpart of the GRB was discovered and identification of NGC 4993
as its host was reported by \citet{2017Sci...358.1556C}.

On the outcome of the merging process, \citet{Granot2017} gave a comprehensive discussion.
Among the potential remnants, a massive NS and the long lasting, differentially rotating supra-massive neutron star (SMNS) do not lead to the formation of a BH-torus system that powers the bursts \citep{Shibata2000, Margalit2015}.
However, a hyper-massive neutron star (HMNS) may also form a BH-torus engine. In case of the GW 170817 event, a  delayed collapse of a HMNS might be the reason for the observed time delay between the merger and GRB phenomena \citep{Granot2017}.
In the same event, the early optical and infrared emission due to r-process is
also in favor of the HMNS scenario \citep{MargalitMetgzer2017}.
Such emission has been confirmed by a number of ground based observations,
e.g., \citet{2017Natur.551...75S}.

On the theoretical ground, it was proposed already e.g. by \citet{1998ApJ...507L..59L} that compact binary mergers eject a small fraction of matter with a subrelativistic velocities. This medium condenses into neutron-rich nuclei, most of which are radioactive and provide a long-term heat source for the expanding envelope.
The first tentative signal of this kind was reported in 2013, when
the ground-based optical and Hubble Space Telescope optical and near-IR
observations of the short-hard GRB 130603B revealed the presence of near-IR emission. It was explained as  the effect of an r-process powered transient \citep{2013ApJ...774L..23B, 2013Natur.500..547T}.

This emission may originate from the long tidal tails of coalescing compact
binaries, as proposed by the hydrodynamic and nuclear reaction network calculations (see, e.g., \citet{2011ApJ...736L..21R}).
In addition,  the r-process  nucleosynthesis may be contributed by
the black hole accretion
disk outflows. Such studies were performed in the past, e.g. in the frame of semi-analytic, time-dependent evolution models \citep{2004ApJ...614..847F, 2012ApJ...746..180W}, as well as in numerical simulations in the pseudo-Newtonian gravity \citep{2015MNRAS.448..541J, 2016MNRAS.463.2323W}.

In the present work, we present a numerical simulation of the short GRB
central engine as composed of the 
rotating black hole, surrounded by a remnant torus which has already formed after the NSNS binary coalescence. We use the full general relativistic
framework and we present an axisymmetric simulation
in the fixed Kerr background metric.
The equation of state (EOS) of the dense matter in the torus
 describes the Fermi gas where the gas pressure is
contributed by partially degenerate nucleons, electron positron pairs, and Helium nuclei. The weak interactions are controlled by the nuclear
equilibrium conditions
and establish the neutronisation level in the torus plane, as well as in the
outflows launched from its surface.
The torus matter is magnetized and its 
accurate evolution is followed when the MRI turbulence
\citep{Balbus_Hawley_1991} drives the mass inflow.
In addition, the magnetic field is responsible for launching 
the uncollimated outflows of the plasma, while the rotation of the black hole
powers the polar jets.
Our simulations include also the neutrino emission, as in the framework of the
so-called neutrino-dominated accretion flows, NDAF \citep{Kohri2005}.

Our numerical scheme is based on the
GRMHD code \textit{HARM} \citep{Gammie_2003, Noble_et_all_2006},
but equipped
with the non-adiabatic EOS. It is embedded in the relativistic conservative
scheme and
interpolation over the temperatures and densities spans several orders of magnitude, as first proposed by \citet{Janiuk2017}.
The main focus, and novelty of the present study is the imprint of the torus properties
on the amount and chemical composition of the emerging ejecta.
We follow the wind outflow, and compute the synthesis of subsequent
r-process elements
on the trajectories, where mass is ejected in sub-relativistic particles.
The effectiveness of the adopted method is that it distributes tracers uniformly in rest-mass density in flow, which is a non-trivial task in numerical relativity \citep{2017CQGra..34u5005B}.

 The current paper is an extension of our previous works \citep{Janiuk2013, Janiuk2017}, where we studied the neutrino cooling and microphysical properties of the GRB central engine.
 In the former studies, the accretion was also modeled with general relativistic MHD simulations, but we did not follow the dynamics of the outflows, and
 we implemented only the nuclear reaction networks producing heavy ions under the nuclear statistical equilibrium \citep{2006NuPhA.777..188H}.
  In the current approach, we follow the neutron rich ejecta, where the
 synthesis of isotopes proceeds faster than the equilibrium timescale, and
 we are able to
 obtain the abundance patters reaching the third peak ($A \sim 195$).
 The effective postprocessing is made with the modular reaction
 network library \citep{2017ApJS..233...18L}.
 
 The article is organized as follows. In Section 2 we present the simulation scheme and the initial configuration of our model.
 We also describe the properties of the EOS (Section 2.2), and the concept of tracer particles (Section 2.3).
  The results showing the general properties of the flow, and the
 nuclear reaction network simulation outcome, appear in Section 3. Discussion and
 conclusions are the subject of Section 4.

\section{The Simulations Setup}

We use the general relativistic magneto-hydrodynamic code, \textit{HARM} \citep{Gammie_2003, Noble_et_all_2006}, to integrate our model under a fixed Kerr metric, i.e. neglecting effects like the self gravity of the  disrupted material, or the BH spin changes.
The {\it HARM} code is a finite volume, shock capturing scheme that solves 
the hyperbolic system of the partial differential equations of GR MHD.
The numerical scheme is based on GR MHD equations with
the plasma energy-momentum tensor, $T^{\mu\nu}$, with contributions from gas and electromagnetic field
\begin{eqnarray*}
{T_{\left(m\right)}}^{\mu\nu}= \rho \enth u^\mu u^\nu + p g^{\mu\nu} \\
{T_{\left(em\right)}}^{\mu\nu}=b^\kappa b_\kappa u^\mu u^\nu+\frac{1}{2} b^\kappa b_\kappa g^{\mu\nu} - b^\mu b^\nu\\
T^{\mu\nu}={T_{\left(m\right)}}^{\mu\nu}+{T_{\left(em\right)}}^{\mu\nu}
\end{eqnarray*}
where $u^{\mu}$ is the four-velocity of gas, $u$ denotes internal energy density, p is pressure,  $b^{\mu}$ is magnetic four-vector, and
$\enth$ is the fluid specific enthalpy, $\enth=(\rho+p+u)/\rho$.
The continuity and momentum conservation equation reads:
\begin{eqnarray*}
(\rho u_{\mu})_{;\nu} = 0 \\
T^{\mu}_{\nu;\mu} = 0.
\end{eqnarray*}
They are brought in conservative form, by implementing the Harten, Lax, van Leer (HLL) solver to calculate numerically the corresponding fluxes. 

In terms of the Boyer-Lindquist coordinates, $\left(r,\theta,\phi\right)$, the black hole is located at $0<r \leq r_{\rm h}$, where $r_{\rm h} = \left(1+\sqrt{1-a^{2}}\right) r_{\rm g}$ is the horizon radius of a rotating black hole with mass $M$ and angular momentum $J$ in geometrized units, $r_{\rm g}=G M /c^2$, and $a$ is the dimensionless Kerr parameter, $a=J/(Mc), 0 \leq a \leq 1$. 
In our simulations we investigate the rotating black holes, $a=0.6-0.9$.

The {\it HARM} code doesn't perform the integration in the Boyer-Lindquist coordinates, but instead in the so called Modified Kerr-Schild ones: $\left(t,x^{(1)},x^{(2)},\phi\right)$ \citep{Noble_et_all_2006}. The transformation between the coordinate systems is given by:

\begin{eqnarray*}
r= R_0 + \exp\left[{x^{(1)}}\right]
\\
\theta = \frac{\pi}{2} \left(1 + x^{(2)}\right) + \frac{1 - \hslope}{2} \sin\left[\pi\left(1 +x^{(2)}\right)\right]
\end{eqnarray*}
where $R_0$ is the innermost radial distance of the grid, $0 \leq x^{(2)} \leq 1$, and $\hslope$ is a parameter that determines the concentration of points at the mid-plane. In our models we use $h=0.3$ (notice that for $h=1$ and a uniform grid on $x^{(2)}$ we obtain an equally spaced grid on $\theta$, while for $\hslope=1$ the points concentrate on the mid plane). The exponential resolution in the $r$-direction leads to higher resolution and it is adjusted to resolve the initial propagation of the outflow. 
Our grid resolution is $256 \times 256$.

\subsection{The Torus Initial Configuration}

The  accreting material is modeled following \citet{Fishbone_Moncrief_1976_ApJ} (hereafter FM) who provided an analytic solution of a constant specific 
angular momentum, in a steady state configuration of a pressure-supported ideal fluid in the Kerr black hole potential.
Other similar configurations like \citet{Chakrabarti1985} with a power law radial evolution of the angular momentum or of independently varying the Bernoulli parameter (sum of the kinetic and potential energy, and enthalpy of the gas) and disk thickness  are also possible \citep{Pennaetal2013}.

In the FM model, the position of the material reservoir is determined by the radial distance of the innermost cusp of the torus, $r_{\rm in}$, and the distance where the maximum pressure occurs, $r_{\rm max}$.
Because of its geometry, the relative difference of the two radii determines also the dimension of the torus, with higher differences resulting to extended cross section. Subsequently the $r_{\rm in}$ and $r_{\rm max}$ determine also the angular momentum value and the distribution of the angular velocity along the torus.

The initial torus is embedded in a poloidal magnetic field, prescribed with the vector potential of 
\begin{equation}
A_{\varphi}={\bar{\rho} \over \rho_{\rm max}} - \rho_{0}
\end{equation}
where $\bar{\rho}$ is the average density in the torus, $\rho_{\rm max}$ is the density maximum, and we use $\rho_{0}=0.2$.
As a consequence, the magnetized flows considered here are not in equilibrium, and the angular momentum is transported. 
Regardless of the specific magnitude of the plasma $\beta-$parameter 
the flow slowly relaxes its initial configuration, becomes geometrically thinner, and launches the outflows.

We examine two sets of models with different initial $\beta-$parameter, defined as the ratio of the fluid's thermal to the magnetic pressure, $\beta \equiv p_g / p_{\rm mag}$ for the torus configurations, 
while every set includes models differing with the black hole spin.

Following \citet{Janiuk2013}, we use the value of the total initial mass of the torus to scale the density over the integration space, and
we base our simulations on the physical units. We use
\begin{eqnarray*}
L_{\rm unit}=\frac{G M}{c^2}=1.48\cdot 10^5 \frac{M}{M_\odot} ~~ \rm{cm} \\
T_{\rm unit}=\frac{r_g}{c}=4.9\cdot 10^{-6} \frac{M}{M_\odot} ~~ \rm{s} \\
\end{eqnarray*}
as the spatial and time units, respectively.
Now, the density scale is related to the spatial unit as $D_{\rm unit}=M_{\rm scale}/L_{\rm unit}^{3}$. We conveniently adopt the scaling factor of 
$M_{\rm scale}=1.5\times 10^{-5} M_{\odot}$, so that the total mass contained 
within the simulation volume (mainly in the FM torus) is about 0.1 $M_{\odot}$, for  a black hole mass of 3 $M_{\odot}$. 
The $r_{\rm in}$ and $r_{\rm max}$ radii of the torus are
having the fiducial values of $r_{\rm in} = (3.5 - 4.5)\,r_g$ and $r_{\rm max} = (9-11)\,r_g$ (see Table \ref{tab:in}).

To sum up, the GRB engines are modeled with the following global parameters: mass of the black hole $M_{\rm BH}$, the torus mass, and black hole dimensionless spin, $a$.  Typical parameters are $M_{\rm BH}=3 \,M_{\odot}$, and $a=0.6-0.9$, while the torus mass is about 0.1 $M_{\odot}$,
as resulting from its size in geometrical units, and adopted density scaling.
The accretion rate onto the black hole, measured as the mass flux transported
through the horizon, is varying with time. Its mean value expressed in
physical units is on the order of 0.1~$M_{\odot}$s$^{-1}$, as converted from the density scaling and time unit.

\subsection{Equation of state and neutrino cooling}

We consider the torus composed of free protons, neutrons, electron-positron pairs, and Helium nuclei.
For a given baryon number density and temperature, the equilibrium condition
is assumed between the reactions of electron-positron capture on nucleons, and neutron decays \citep{1998PhRvD..58a3009R}.

    Namely, the ratio of free protons, is determined from the equilibrium between the transition reactions from neutrons to protons, and from protons to neutrons: $p + e^{-} \rightarrow n + \nu_e$, $p+\bar\nu_e \rightarrow n+e^{+}$,  $p+e^{-}+\bar\nu_e \rightarrow n$,
    $n+e^{+}\rightarrow p+\bar\nu_e$, $n + \nu_e \rightarrow p + e^{-}$, and $n\rightarrow p+e^{-}+\bar\nu_e$. The balance equation has a form:

  \begin{equation}
n_{\rm p}(\Gamma_{p + e^{-} \rightarrow n + \nu_e} +
\Gamma_{p+\bar\nu_e \rightarrow n+e^{+} } +
\Gamma_{ p+e^{-}+\bar\nu_e \rightarrow n}) =
n_{\rm n}(
\Gamma_{n+e^{+}\rightarrow p+\bar\nu_e} +
\Gamma_{n + \nu_e \rightarrow p + e^{-}} +
\Gamma_{n\rightarrow p+e^{-}+\bar\nu_e}).
  \end{equation}

The above reaction rates are the sum of forward and backward rates and the
    appropriate formulae for $\Gamma$'s are given in \citet{Kohri2005}.

      The number density of fermions under arbitrary degeneracy is
      determined by
      \begin{equation}
n_{\rm i} = {\sqrt{2}\over \pi^{2}}
{(m_{i}c^{2})^{3} \over (\hbar c)^{3}}\beta_{i}^{3/2}
\left[F_{1/2}(\eta_{\rm i},\beta_{\rm i})+{1\over 2} \beta_{\rm i}F_{3/2}(\eta_{\rm i},\beta_{\rm i})\right]
        \end{equation}
      with $F_{\rm k}$ being the Fermi-Dirac integrals of the order $k$, and $\beta_{i}=kT_{i}/(m_{i}c^{2})$.

  This is supplemented with the charge neutrality condition,
  $n_{e^{-}}=n_{e^{+}}+n_{p}+n^{0}$, where $n^{0}=2 n_{\rm He}=(1-X_{\rm nuc}) n_{b}/2$ is the number of protons in Helium, and the conservation of baryon number, $n_{\rm p}+n_{\rm n}= X_{\rm nuc}n_{\rm b}$. The fraction of free nuclei is a strong function of density and temperature, and is adopted from the fitting formula after \citet{Quian1996}.

We can define the proton-to-baryon number density ratio $Y_{\rm p}$, and the electron fraction:
\begin{equation}
  Y_{\rm e} = {n_{\rm e^{-}} - n_{\rm e^{+}} \over n_{\rm b}}.
  \label{eq:yee}
\end{equation}

The free species may have an arbitrary degeneracy level.
The total pressure is contributed by free nucleons, pairs, radiation, alpha particles, and also trapped neutrinos
 \citep{yuan2005, janiuk2007}, and is given by:
\begin{equation}
P = P_{\rm nucl}+P_{\rm He}+P_{\rm rad}+P_{\nu}
\end{equation}
where  {\bf $P_{\rm nucl}=P_{e^{-}} + P_{e^{+}}+P_{n}+P_{p}$ } and 
\begin{equation}
P_{\rm i} = {2 \sqrt{2}\over 3\pi^{2}}
{(m_{i}c^{2})^{4} \over (\hbar c)^{3}}\beta_{i}^{5/2}
\left[F_{3/2}(\eta_{\rm i},\beta_{\rm i})+{1\over 2} \beta_{\rm i}F_{5/2}(\eta_{\rm i},\beta_{\rm i})\right]
\label{eq:pressure}
\end{equation}
where $\eta_{\rm e}$, $\eta_{\rm p}$ and $\eta_{\rm n}$ are the reduced chemical
potentials. Here, $\eta_i = \mu_i/kT$ is the degeneracy
parameter (where $\mu_i$ is the standard chemical
potential). Reduced chemical potential of positrons is
$\eta_{\rm e+}=-\eta_{\rm e}-2/\beta_{\rm e}$.

The total pressure of subnuclear matter is mainly contributed by electrons,
and therefore it is influenced by the changes in the electron fraction.
Reduced electron chemical potentials are somewhat larger than unity, because at these accretion rates which we consider here, electrons are slightly degenerate.
The pressure of Helium is taken to be of ideal gas, $P_{\rm He}=(n_{b}/4) (1-X_{\rm nuc}) kT$. The radiation pressure, $P_{\rm rad}=(4\sigma)/(3c)T^{4}$, is in GRB disks a few orders of magnitude smaller than other components.

\subsection{Neutrino cooling}

The reactions of the electron and positron capture on nucleons (a.k.a. URCA reactions, see above), and also
the electron-positron pair annihillation, $e^{+}+e^{-} \rightarrow \nu_{i}+\bar\nu_{i}$, nucleon bremsstrahlung,  $n+n \rightarrow n +n + \nu_{i}+\bar\nu_{i}$,  and plasmon decay, $\tilde\gamma \rightarrow \nu_{e}+\bar\nu_{e}$, are
producing neutrinos, which act as a cooling mechanism for the plasma.

  The cooling rates due to the bremsstrahlung and plasmon decay, are $q_{\rm brems}=3.35\cdot 10^{27}\rho_{10}^{2}T_{11}^{5.5}$, and
  $q_{\rm plasm}= 1.5\cdot 10^{32}T_{11}^{9}\gamma_{p}^{6}e^{-\gamma_{p}}(1+\gamma_{p})(2+\gamma_{p}^{2}/(1+\gamma_{p})$ where
  $\gamma_{p}=5.56\cdot 10^{-2}\sqrt{(\pi^{2}+3\eta_{e}^{2})/3}$
  \citep{Ruffert1996}.

  The cooling rates due to  URCA processes, $q_{\rm urca}$, and pair annihillation, $q_{\rm pair}$, reactions
  are having more complex forms, and can be found i.e. in \citet{janiuk2007}
  (see equations A7-A15 therein). They involve the distribution functions, and the blocking factors, which describe the extent
  on which neutrinos are trapped (see below).
  In the GRB accretion disk, we consider the neutrino transparent and opaque regions, and transition between the two. In terms of the blocking factors, $0\le b_{i}\le 1$, the neutrinos of $\nu_{i}$ flavor might be freely escaping
  or trapped, consistently with the two-stream approximation \citep{2002ApJ...579..706D}.
  The blocking factor of trapped neutrinos is however used only for the URCA emissivities.
  For the annihillation reaction it is not used, because these emissivities are much smaller, and do not change the electron fraction.

  Trapped neutrinos give a contribution to the pressure with a component
  \begin{equation}
P_{\nu} = {7\over 8}{\pi^{2} \over 15}{(kT)^{4} \over 3 (\hbar c)^{3}} \sum_{i=e,\mu,\tau}{0.5 (\tau_{a,i}+\tau_{s}) + {1 \over \sqrt{3}} \over 0.5 (\tau_{a,i}+\tau_{s}) + {1 \over \sqrt{3}} + {1 \over  3\tau_{a,i}}  }
    \end{equation}
  where $\tau_{a,i}$ and $\tau_{s}$ denote absorption and scattering.
   Here $\tau_{a,i}=(H/((7/8)\sigma T^{4})) q_{a, \nu_{i}}$, where $q_{a, \nu_e}=q_{pair}+q_{urca}+q_{plasm}+{1\over 3}q_{brems}$ and
  $q_{a, \nu_\mu}=q_{pair}+{1\over 3}q_{brems}$.
The scale $H$ is the local thickness of the disk given by the pressure scaleheight.
The free escape of neutrinos is further limited by their scattering on neutrons and protons, for which we use a formula
  $\tau_{\rm s} = 24.3\cdot 10^{-5}((kT/m_{e}c^{2}))^{2}H (C_{s,p} n_{p} + C_{s,n} n_{n})$, with $(C_{s,p}=[4(C_{V}-1)^{2}+5 \alpha^{2}]/24$
and $(C_{s,n}=[1 +5 \alpha^{2}]/24$, $C_{V}=1/2+2 \sin^{2}\theta_{C}$, and $\sin^{2}\theta_{\rm C}=0.23$.
The 'blocking factor' for electron neutrino
is then defined as $b_{e} =((\tau_{a,e}+\tau_{s})/2+1/\sqrt{3})/(\tau_{a,e}/2+1/\sqrt{3}+1/(3 \tau_{a,e})$.

    Neutrino cooling rate is finally computed as
    \begin{equation}
Q_{\nu} = {(7/8)\sigma T^{4} \over (3/4)}\sum_{i=e,\mu,\tau} {1 \over {0.5 (\tau_{a,i}+\tau_{s}) + {1 \over \sqrt{3}} + {1 \over  3\tau_{a,i}}}   }
    \end{equation}
    in the optically thick regime.
    In the optically thin regime, it will be given by $Q_{\nu} = H(q_{\rm pair}+q_{\rm urca}+q_{\rm plasm}+q_{\rm brems})$.

We store the neutrino cooling rate as a function of the rest mass density and temperature,
and we also store the corresponding electron fraction and total pressure values,
 that result from the EOS.
The numerically computed EOS is tabulated and the pressure that is used in the MHD evolution, comes out from
interpolation over the density and temperature over the grid,
by means of the spline method \citep{Akima1970}.
The GR MHD 
computation  in this case is non-trivial,
and also technically demanding,
as the numerical scheme  solves at every time-step for the
inversion between the so called 'primitive' and 'conserved' 
variables, the latter being
the total energy and comoving density (see e.g., \citet{Noble_et_all_2006}).
Our modified HARM scheme,
introduced previously in \citet{Janiuk2017}, takes into account
the total pressure as given by Eq. \ref{eq:pressure},
and also its derivatives over density and energy, when computing the sound speed square.
Hence the magneto-sonic velocity, defined as in \citep{2015CQGra..32i5007I} 
is then properly adopted by the MHD stress tensor and source
terms in the equations of motion.
Notice that in the pioneering works dealing with GR MHD disks \citep{McKinney_2012}, the adiabatic equation of state was used in the form of
  $p=(\gamma_{\rm ad}-1)u$ (where $\gamma_{\rm ad}$ is an adiabatic index)
during the dynamical simulations because they were addressed to much lower temperature and density systems, such as AGN disks.

\subsection{The use of tracer particles}

We define the tracers already while initializing the GR MHD simulation. In every grid cell we check first, if the density is larger than some minimum value (e.g., $10^{-4}$ of the maximum density in the torus), hence we pick all the particles which are embedded in the densest parts of the accreting flow.
During the evolution, we update the positions of tracers according to their
new (contravariant) velocity components. The linear interpolation of the four-velocities is performed, so that we are always in the grid cell centers.

\begin{figure}
\begin{tabular}{cccc}
  \hspace{-0.9cm}
  \includegraphics[width=0.29\textwidth]{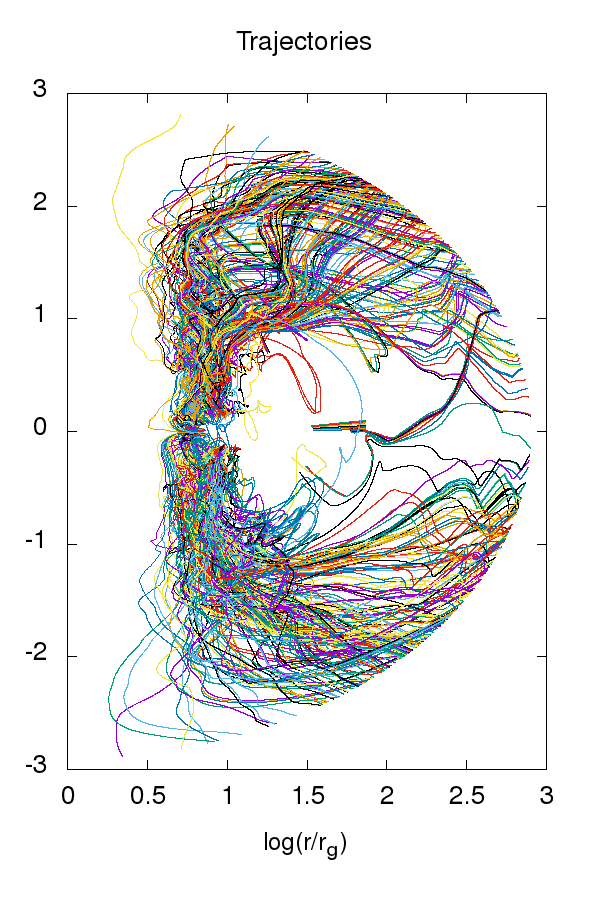} & 
  \hspace{-0.9cm}
  \includegraphics[width=0.29\textwidth]{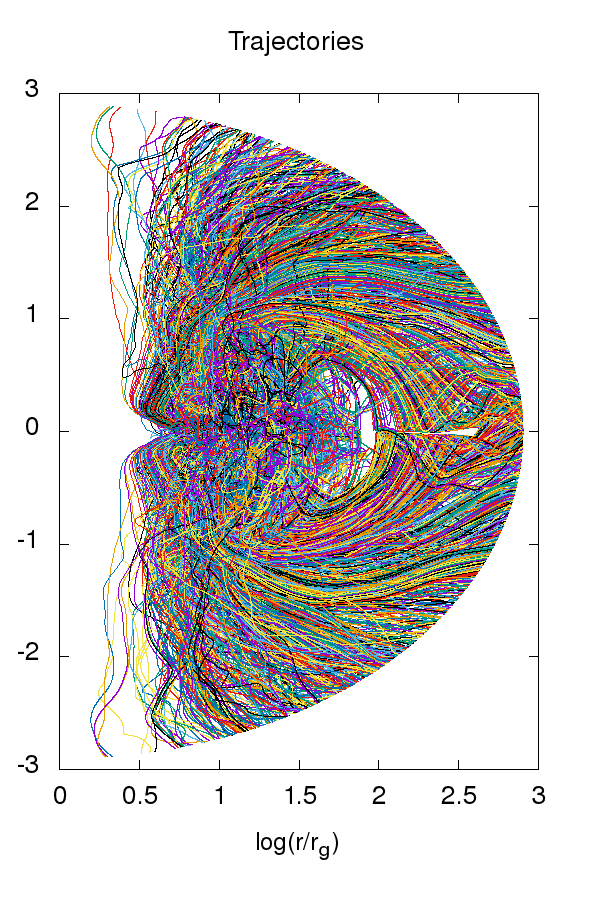}  & 
  \hspace{-0.9cm}
  \includegraphics[width=0.29\textwidth]{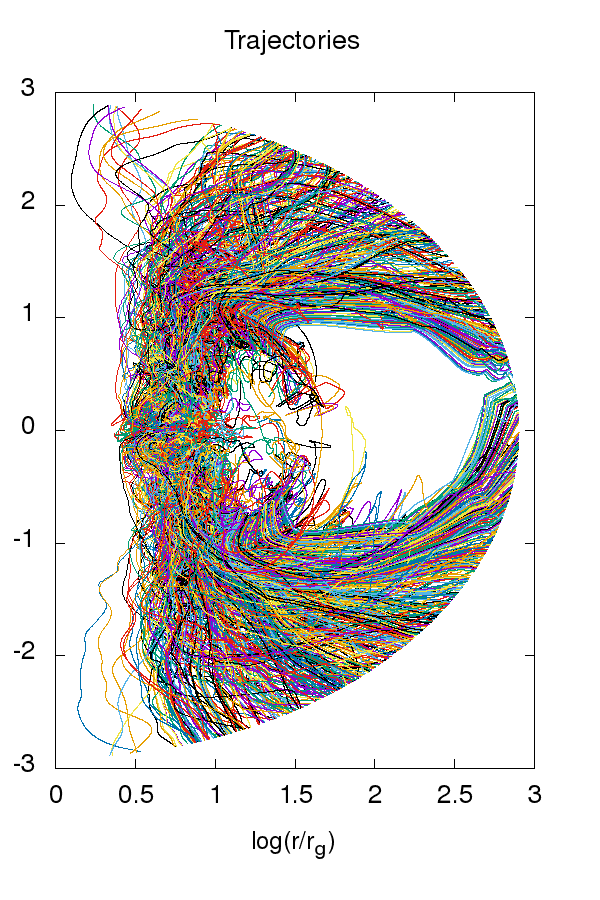} & 
  \hspace{-0.9cm}
  \includegraphics[width=0.29\textwidth]{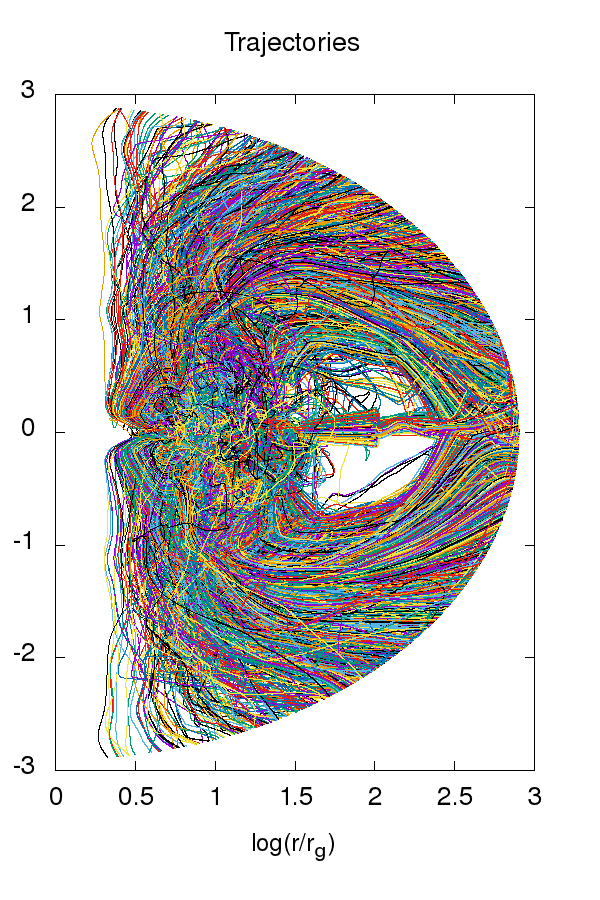}    
\end{tabular}
\caption{Trajectories of the outflowing particles (that reached the outer boundary at the end of the simulation). Parameters of the models are black hole spin $a=0.6$ and $a=0.9$, and plasma magnetisation, $\beta=100$, or $\beta=10$. Models, from left to right, are: LS-Therm, LS-Magn, HS-Therm, HS-Magn. } 
     \label{fig:in_beta}
\end{figure}

While tracking the moving particle, we record their coordinates, density, temperature, and electron fraction, 
which will be used later to compute the nucleosynthesis.
This is done always, whenever the ultimate fate of the trajectory is determined.
The 
particles can escape from the computational domain either through the inner boundary, $R_{\rm in}$, or through the
outer boundary, $R_{\rm out}$. While the inflow corresponds to the black hole accretion, in this study
we are interested mostly in the outflow properties.
Figure \ref{fig:in_beta} shows the trajectories of tracer particles which left through $R_{\rm out}=1000 ~r_{\rm g}$ as computed during the evolution of 
the torus until time $t_{f}=20000 ~M$. We present four models, that are listed in Table \ref{tab:in}.
In Fig. \ref{fig:in_beta}, we plotted only the uncollimated outflows. The collimated ones (i.e. the ``jets''), and ignored. These traces are defined as having the polar angles less than a minimum value (here: $\theta_{\rm min}=0.02 \pi$), as measured from either of the polar semi-axes.

The trajectories recorded in such way are further post-processed to compute the  $r$-process nucleosynthesis. Given the evolution of density, temperature, and electron fraction, they provide input for the thermonuclear reaction network.

\section{Results}

\begin{deluxetable}{lcccccccr}

\tablecaption{Summary of the models \label{tab:in}}

\tablehead{
  \colhead{Model} 
  & \multicolumn{2}{p{3cm}}{\centering  Torus mass \\ $M_{\odot}$ units \\ $M_{0} \qquad\;\;\;\; M_{end}$}
  & \multicolumn{1}{p{2cm}}{\centering BH spin\\ $a$} 
  & \multicolumn{1}{p{2cm}}{\centering $r_{\rm in}$\\ $r_{\rm g}$ units } 
  & \multicolumn{1}{p{2cm}}{\centering $r_{\rm max}$\\ $r_{\rm g}$ units } 
  & \multicolumn{1}{p{2cm}}{\centering $\beta_0$ } 
  & \multicolumn{1}{p{2cm}}{\centering $\dot M_{out}$ \\ $M_{\odot}$ s$^{-1}$ units}
  &  \multicolumn{1}{p{2cm}}{\centering $\Delta M_{out}$\\ $M_{\odot}$ units }
}
\startdata
      HS-Therm &0.1031  & 0.0848 & 0.9 & 3.8 & 9.75 & 100 & $7.64\cdot 10^{-2}$ & $4.42\cdot 10^{-3}$ \\
      HS-Magn & 0.1031  & 0.0741 & 0.9  & 3.8 & 9.75 & 10 & $8.26\cdot 10^{-2}$ & $1.72\cdot 10^{-2}$ \\
      LS-Therm & 0.1104  & 0.0895 & 0.6 & 4.5 & 9.1 & 100 & $8.76\cdot 10^{-2}$ & $2.29\cdot 10^{-3}$ \\
      LS-Magn & 0.1104   & 0.0682 & 0.6  & 4.5 & 9.1 & 10 & $1.20\cdot 10^{-1}$ & $1.81\cdot 10^{-2}$ \\
\enddata
\tablecomments{The first two models refer to a highly spinning black hole in a short GRB engine, and the last two models represent moderately spinning black hole.
  The inner radius of the torus, $r_{\rm in}$, the radius of pressure maximum, $r_{\rm max}$, and the plasma-$\beta$ are given as the initial state parameters.
  The last two columns give the mass loss rate through the outer boundary, averaged over
  the simulation time, and the cumulative mass lost in the ouflow.
}
\end{deluxetable}

\subsection{General structure of the outflow}

Our simulations are divided into two sets with the \textit{Magn} class of models referring to a weakly magnetized torus and the \textit{Therm} class to torus with significantly (ten times) higher plasma $\beta$-parameter.
Table~\ref{tab:in} presents summary of the models and their initial parameters, the total mass of the disk at initial state, and the averaged mass loss rate through the outer boundary.
The physical conditions in the flow, namely its density, temperature, and electron fraction, are governed by the global parameters: accretion rate, black hole mass and its spin, and the magnetic field normalization.

The overall evolution follows a similar pattern for all models that have been investigated. The FM torus initial state is close to a steady state retaining its structure for a relatively long integration time. Its innermost, densest part
is enclosed within $\sim 50 r_{g}$, and forms a narrow cusp through which the matter sinks under the black hole horizon. The surface layers of the torus, which extend to about 200 $r_{g}$, form a kind of corona. This region is the base of
sub-relativistic outflows that start being launched from the center as soon as the initial condition of the simulation is relaxed (this is at about 2000 M, which is 0.03 seconds for the adopted black hole mass).
The models were evolved until the time $t_{\rm f}=20000$\,M (geometrical units, $M=G M_{\rm BH}/c^{3}$, make $t_{f}=0.3$ s for $M_{\rm BH}=3 M_{\odot}$).

\begin{figure}
    \centering
    \includegraphics[width=0.3\textwidth]{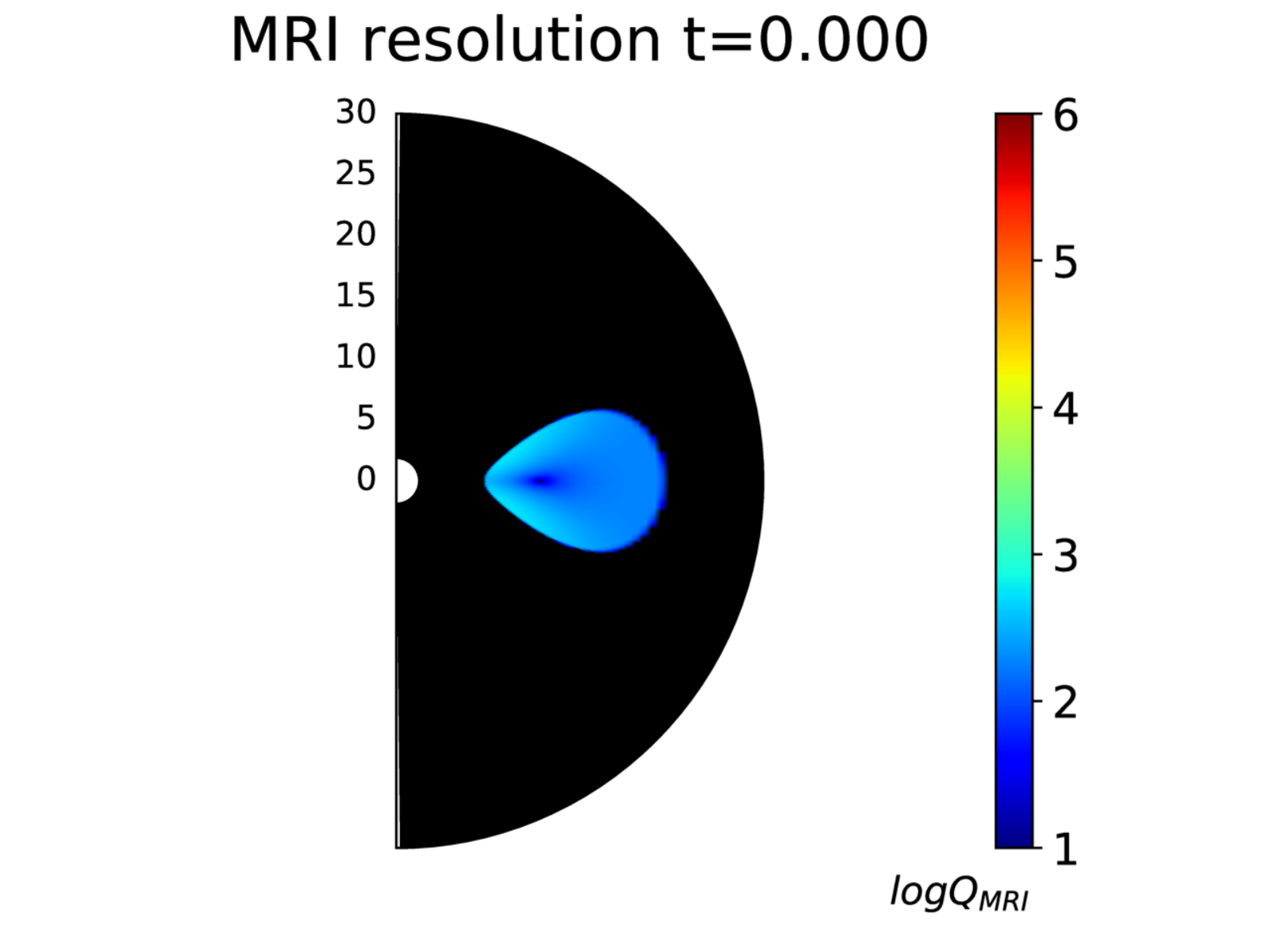}
     \includegraphics[width=0.3\textwidth]{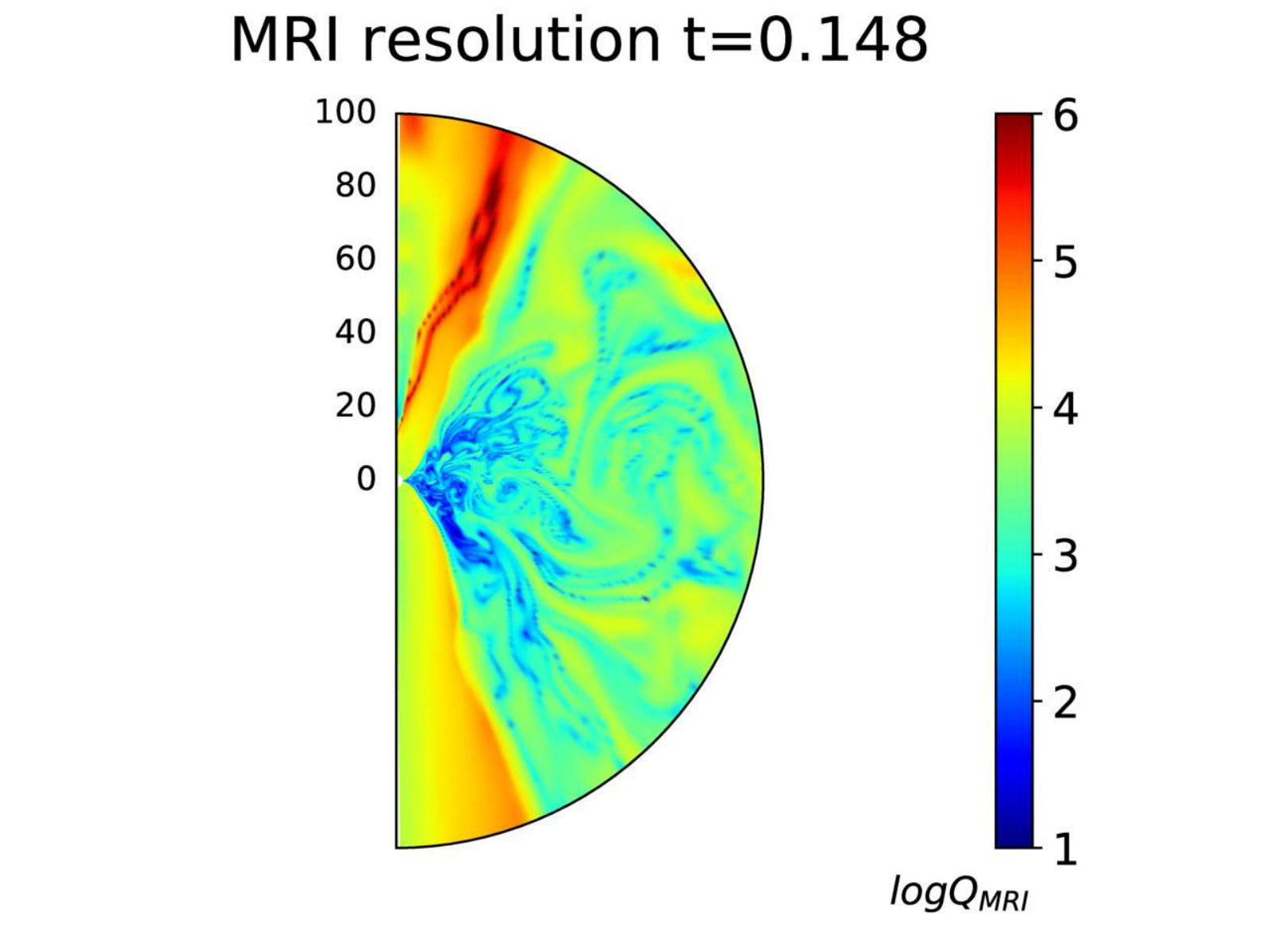}
     \includegraphics[width=0.3\textwidth]{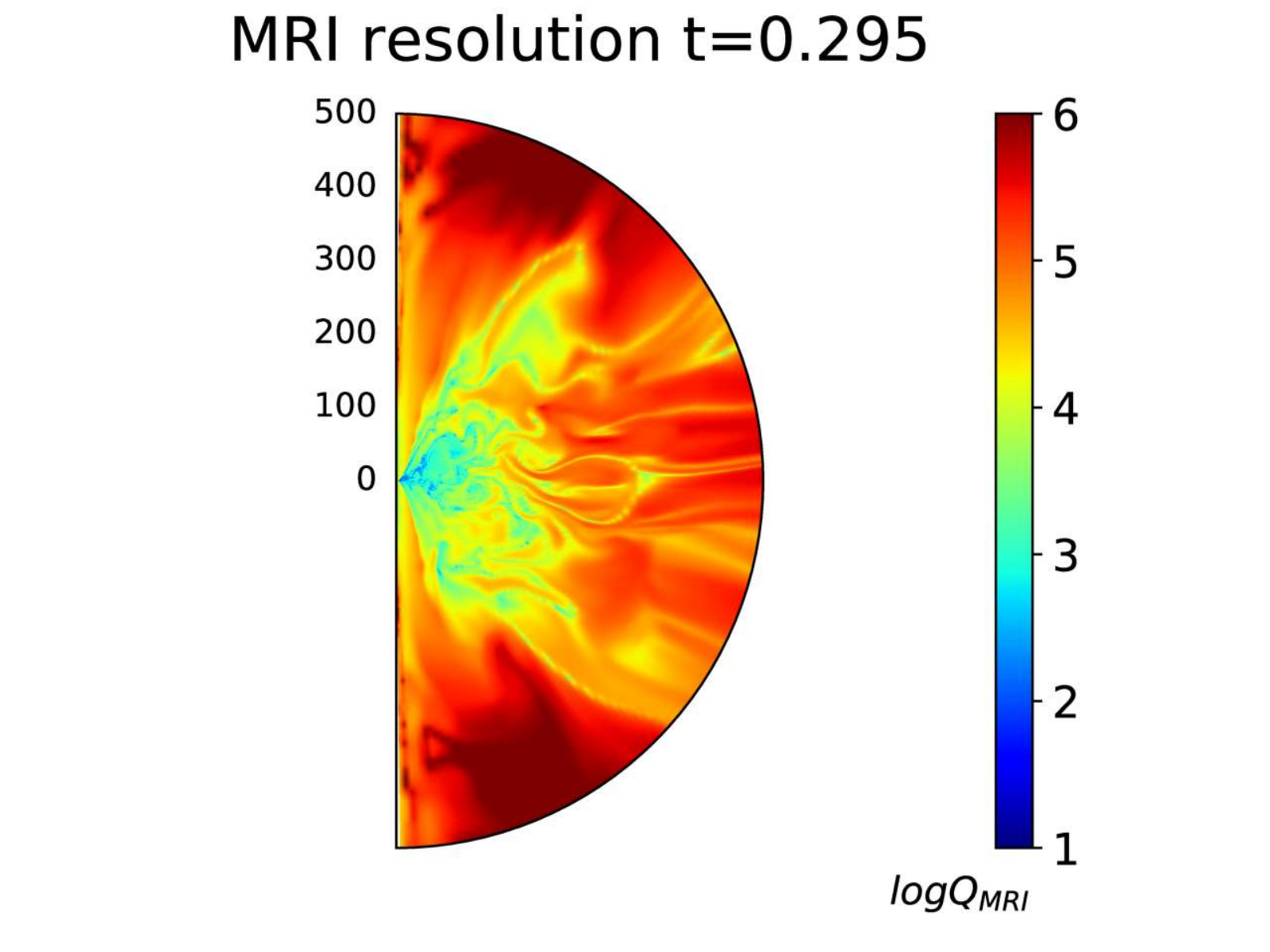}
     \caption{Resolution of the MRI turbulence, in terms of number of grid cells per wavelength of the fastest growing mode. Plots show initial state and
       evolved states at t=0.148 s (10000 M), and t=0.295 (20000 M),
       for the model LS-Magn. Note changing spatial scale (expressed in $r_{\rm g}$ units)}
     \label{fig:qmri}
    \end{figure}

  The magneto-rotational turbulence is resolved in our simulations with the proper scaling of the cell sizes.
  We check this by computing the ratio of the wavelength of the fastest growing mode, as given by:
  \begin{equation}
\lambda_{\rm MRI} = { 2 \pi \over \Omega} {b \over \sqrt {4\pi \rho h + b^{2}}}
  \end{equation}
  We find that our grid provides always at least 10 cells per wavelength, with
  the MRI being at least marginally resolved inside the torus, and well resolved
  in the regions of the outflow.
  In Figure \ref{fig:qmri} we plot the ratio of the $\lambda_{\rm MRI}$
  with respect to the local grid resolution, i.e. $Q_{\rm MRI}={\lambda_{\rm MRI} \over {\Delta\theta}}$ (see e.g. \citet{siegel2018}), for three representative times in simulation LS-Magn.

\begin{figure*}
\begin{tabular}{ccc}
  \hspace{-10mm}\includegraphics[width=0.39\textwidth]{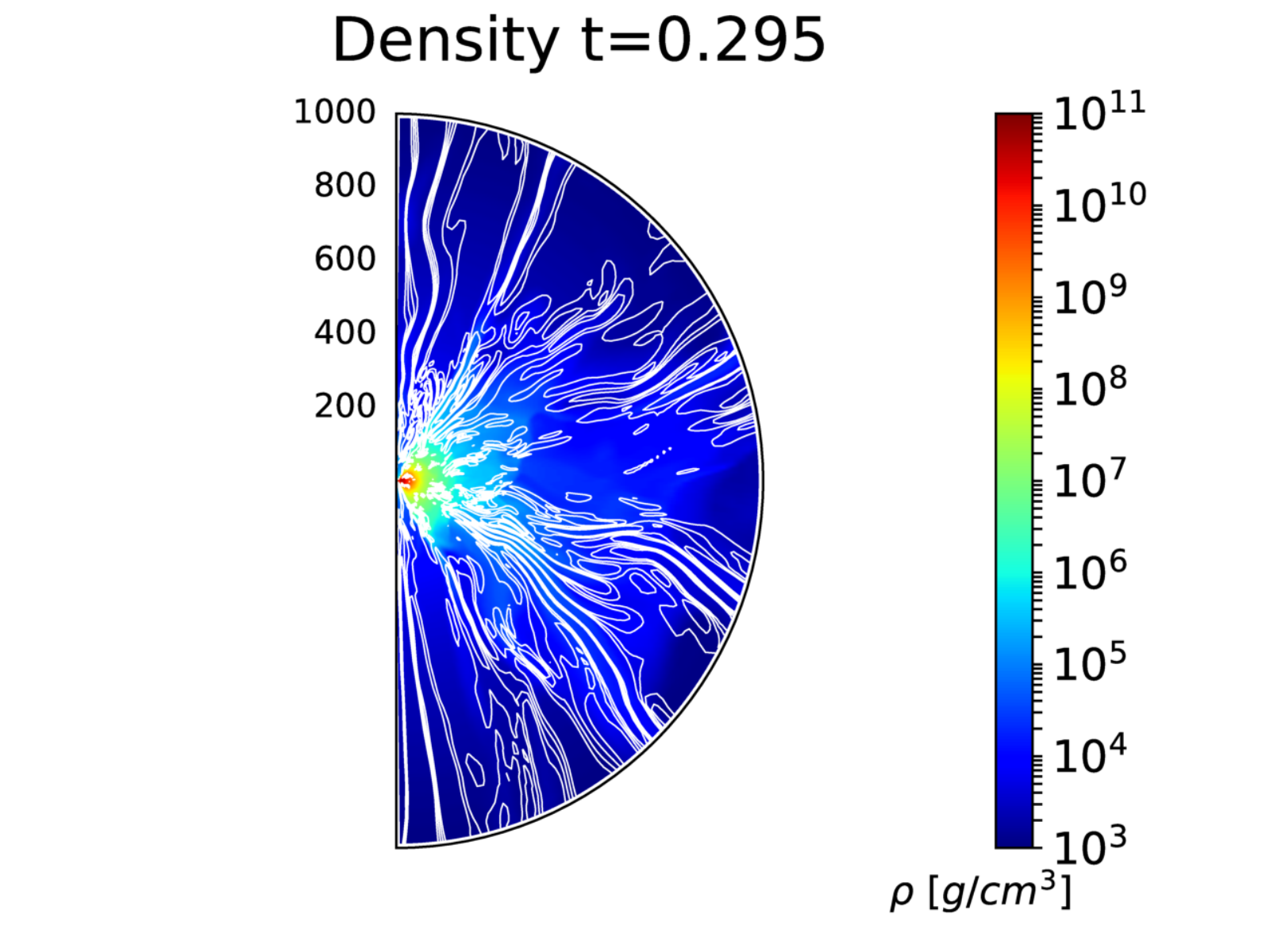} &
  \hspace{-10mm}\includegraphics[width=0.39\textwidth]{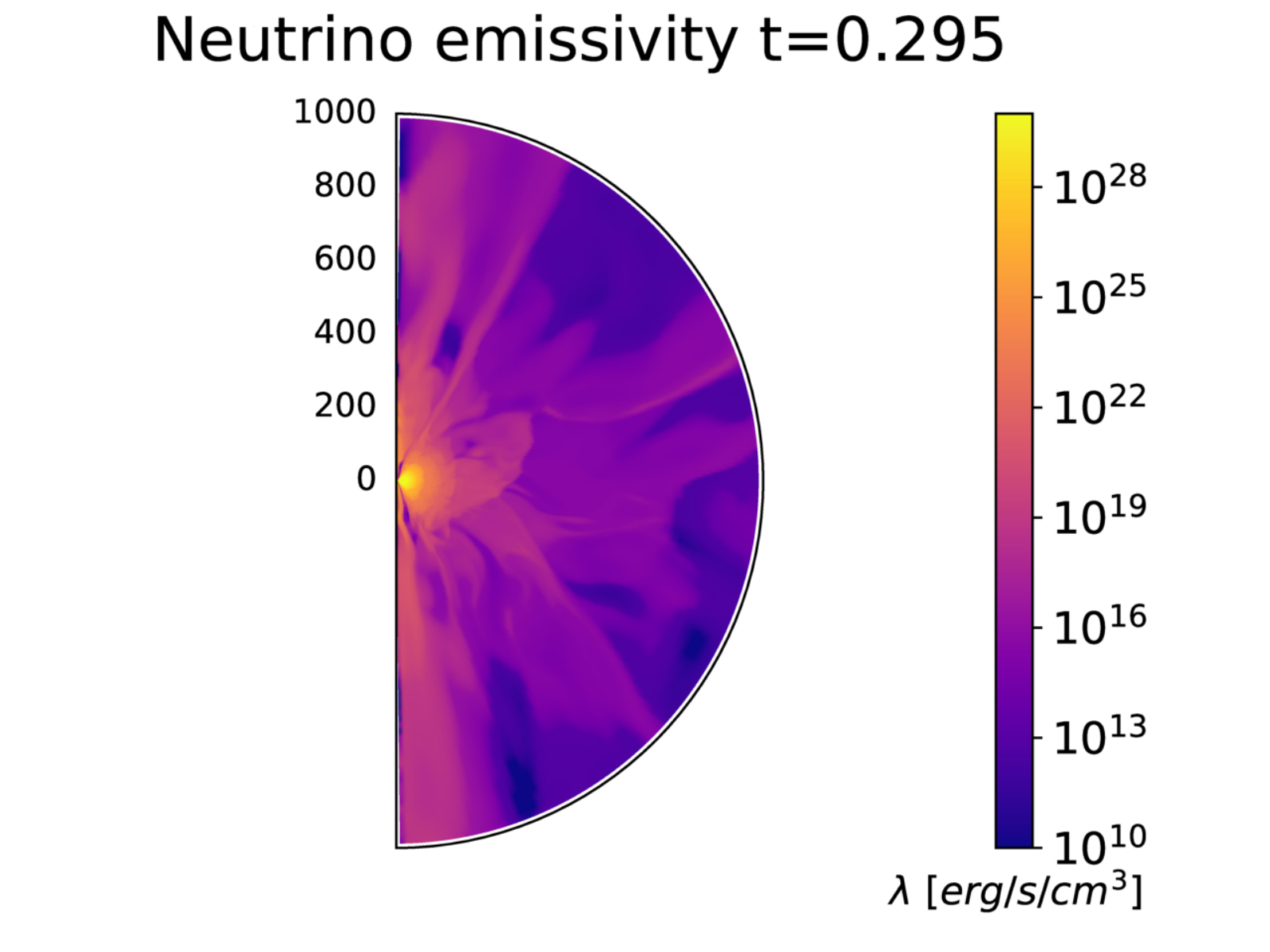} &
  \hspace{-12mm}\includegraphics[width=0.39\textwidth]{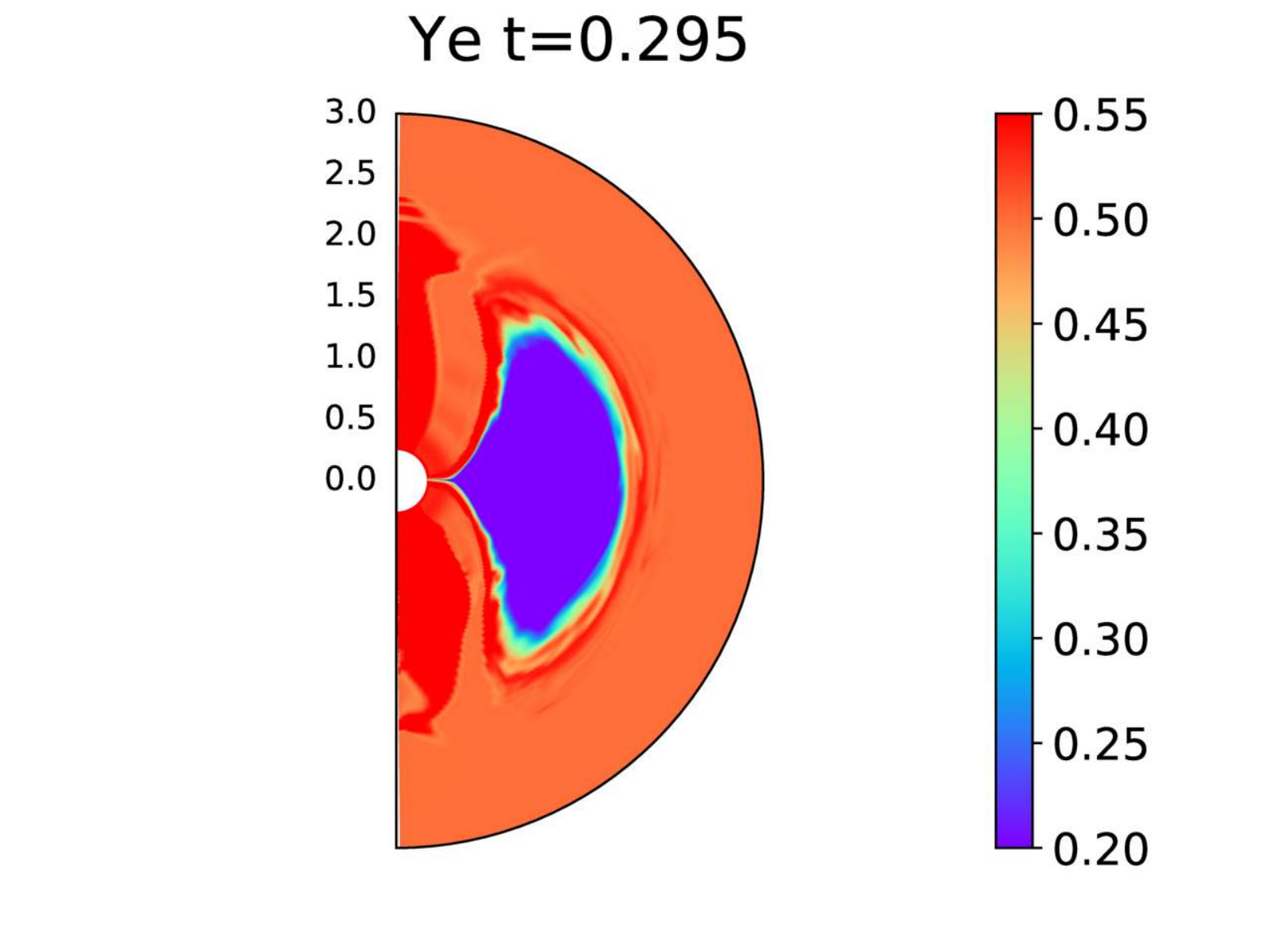} 
\\
  \hspace{-12mm}\includegraphics[width=0.39\textwidth]{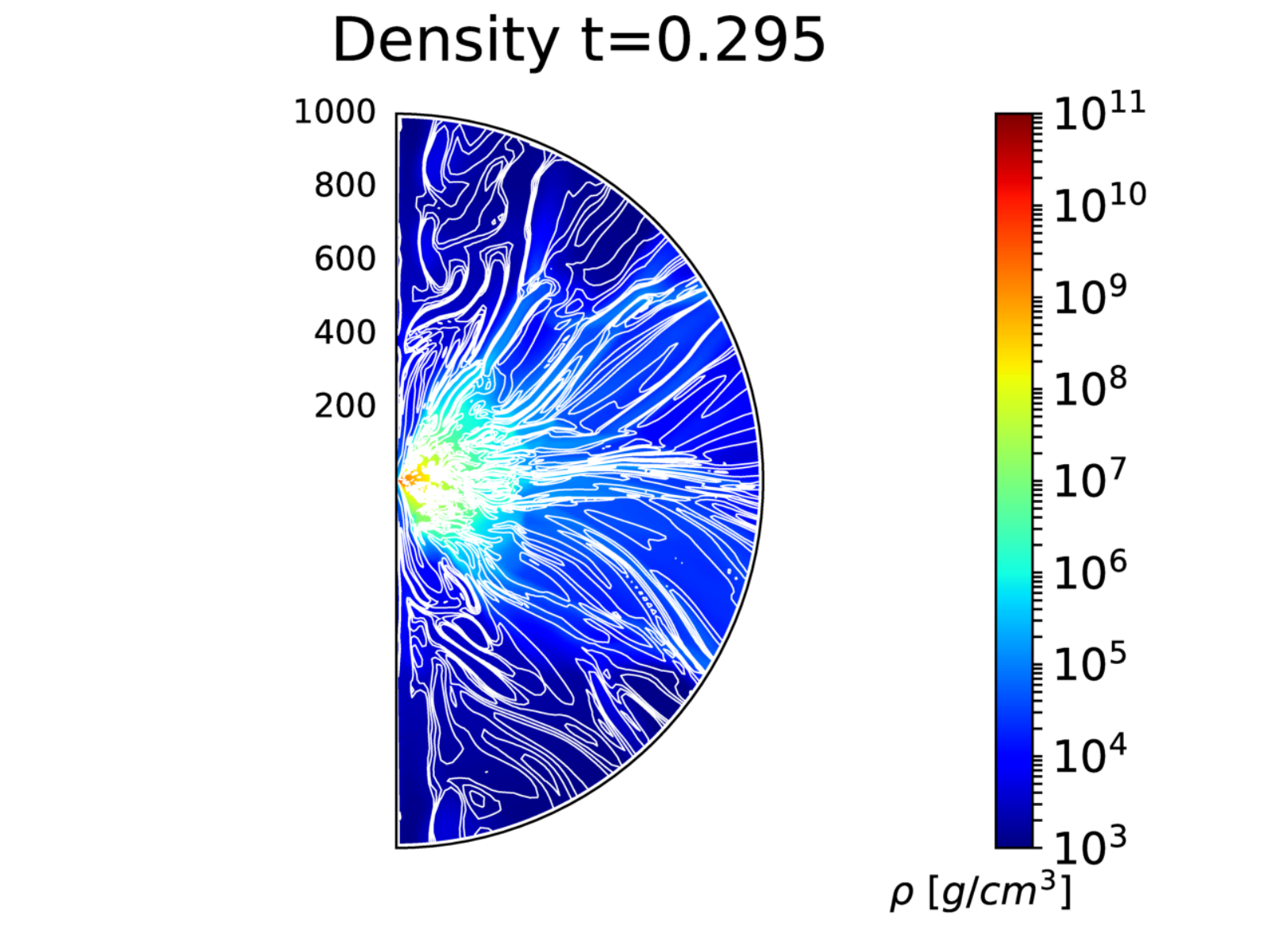}  &
  \hspace{-12mm}\includegraphics[width=0.39\textwidth]{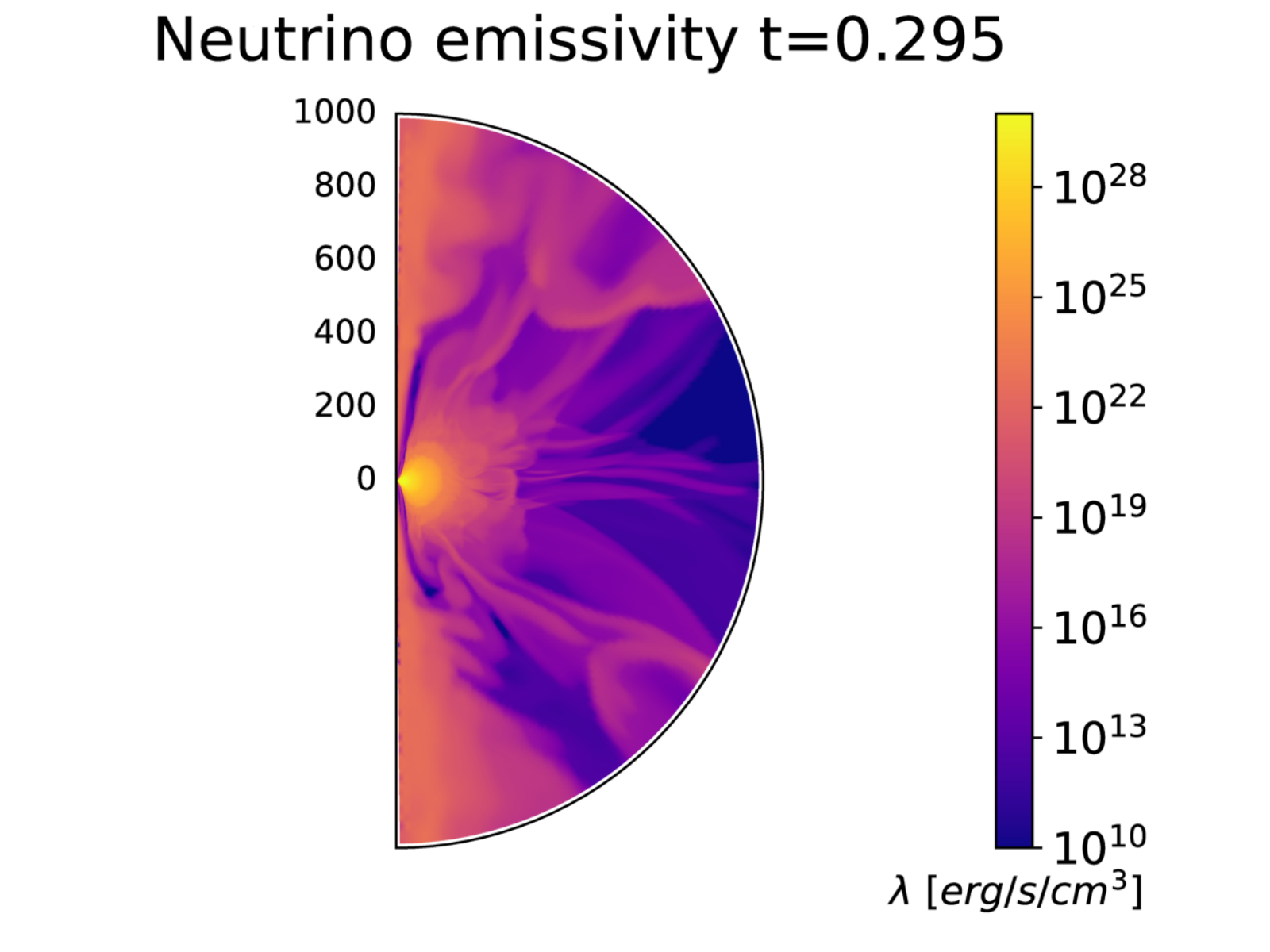}  &
  \hspace{-12mm}\includegraphics[width=0.39\textwidth]{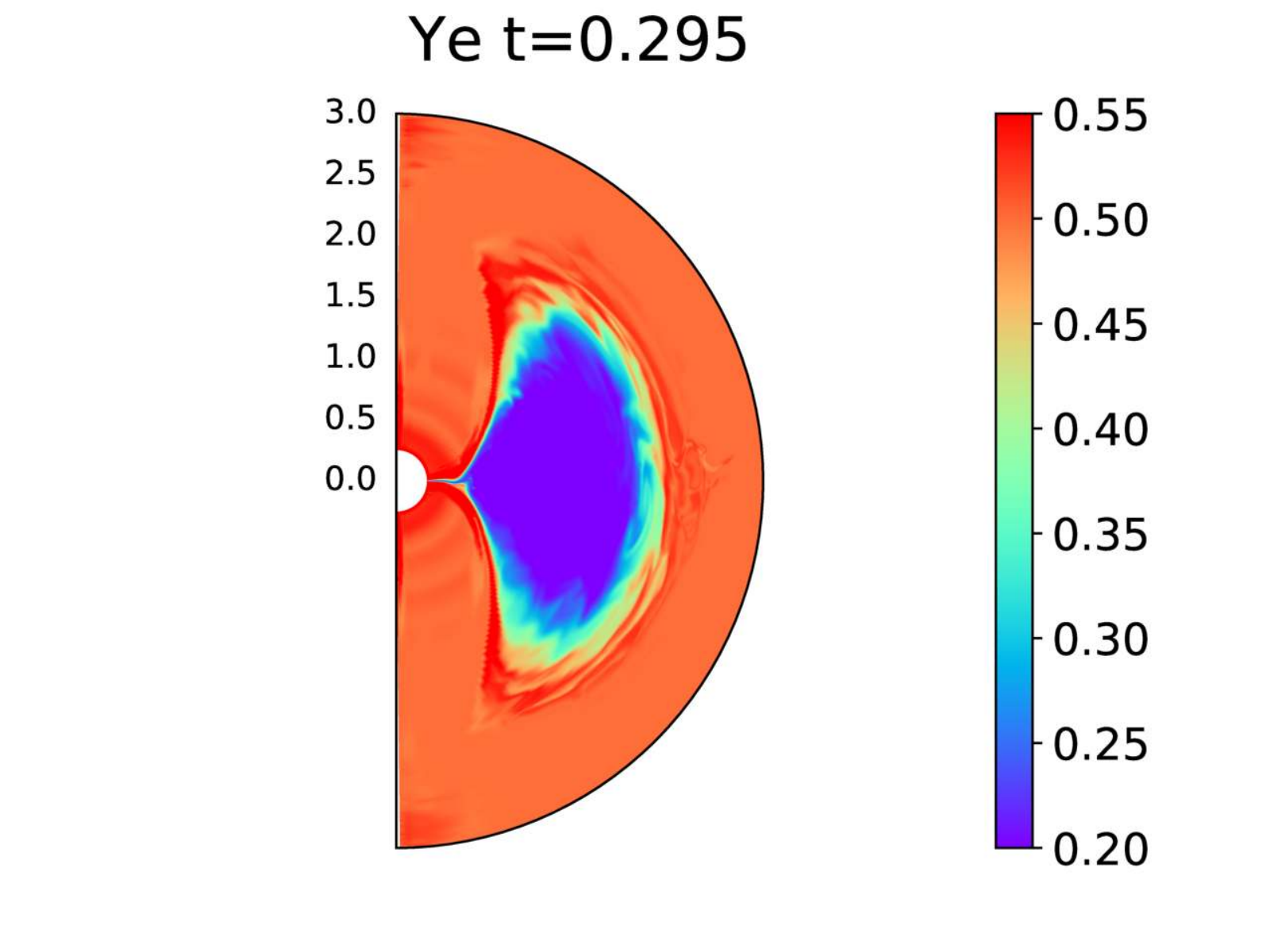}  
\\
  \hspace{-12mm}\includegraphics[width=0.39\textwidth]{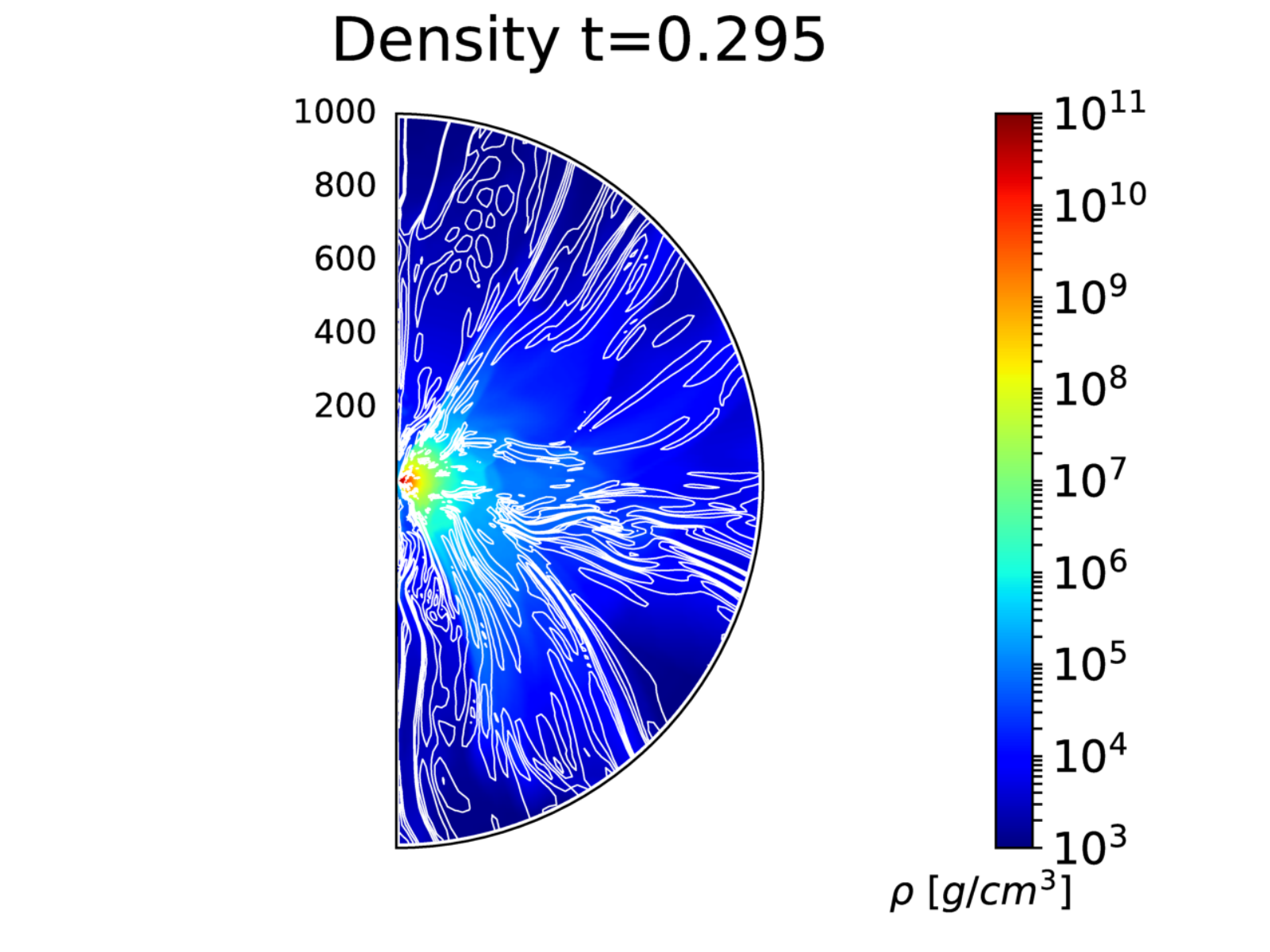} &
  \hspace{-12mm}\includegraphics[width=0.39\textwidth]{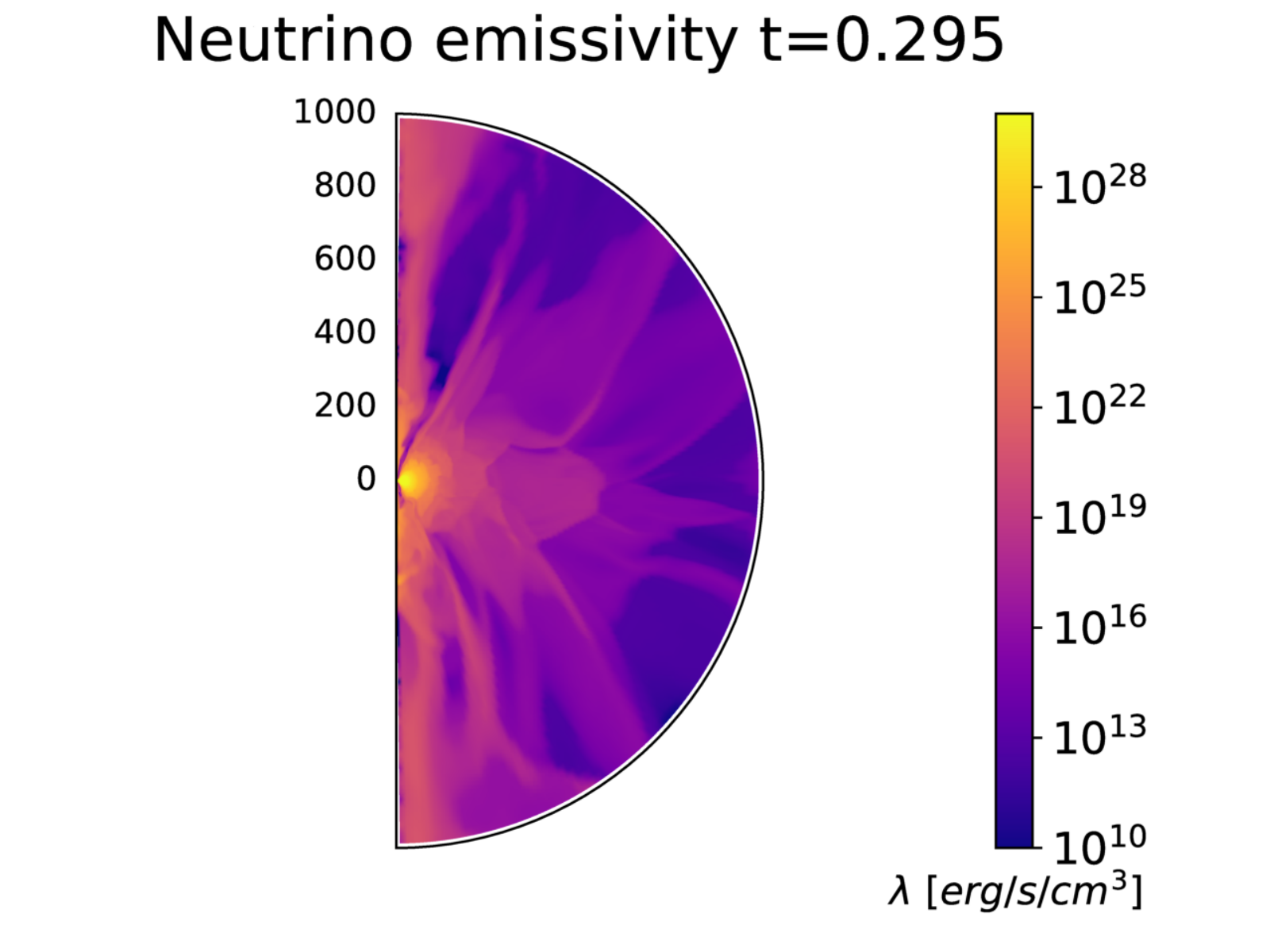} &
  \hspace{-12mm}\includegraphics[width=0.39\textwidth]{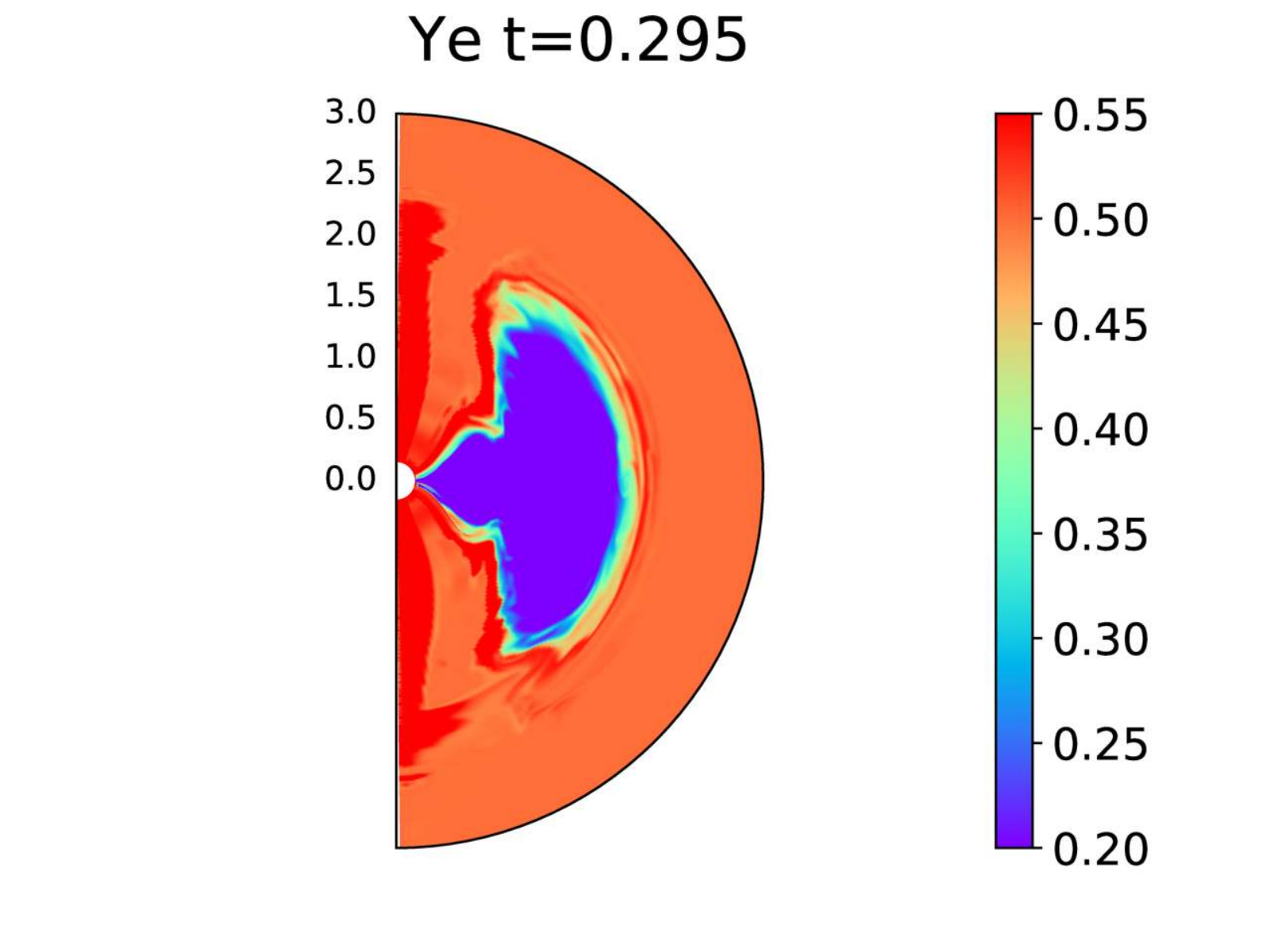} 
\\
  \hspace{-12mm}\includegraphics[width=0.39\textwidth]{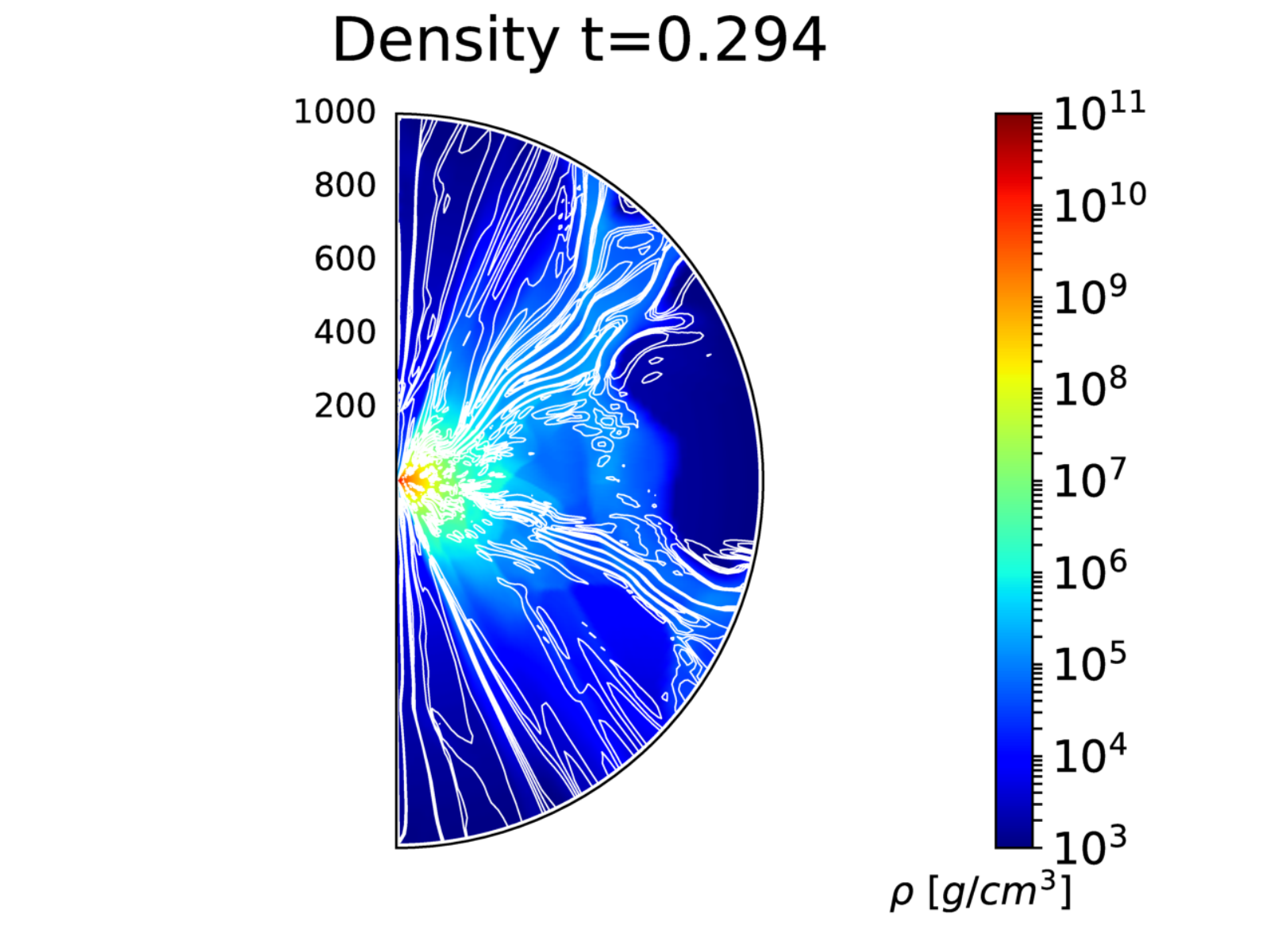} &
  \hspace{-12mm}\includegraphics[width=0.39\textwidth]{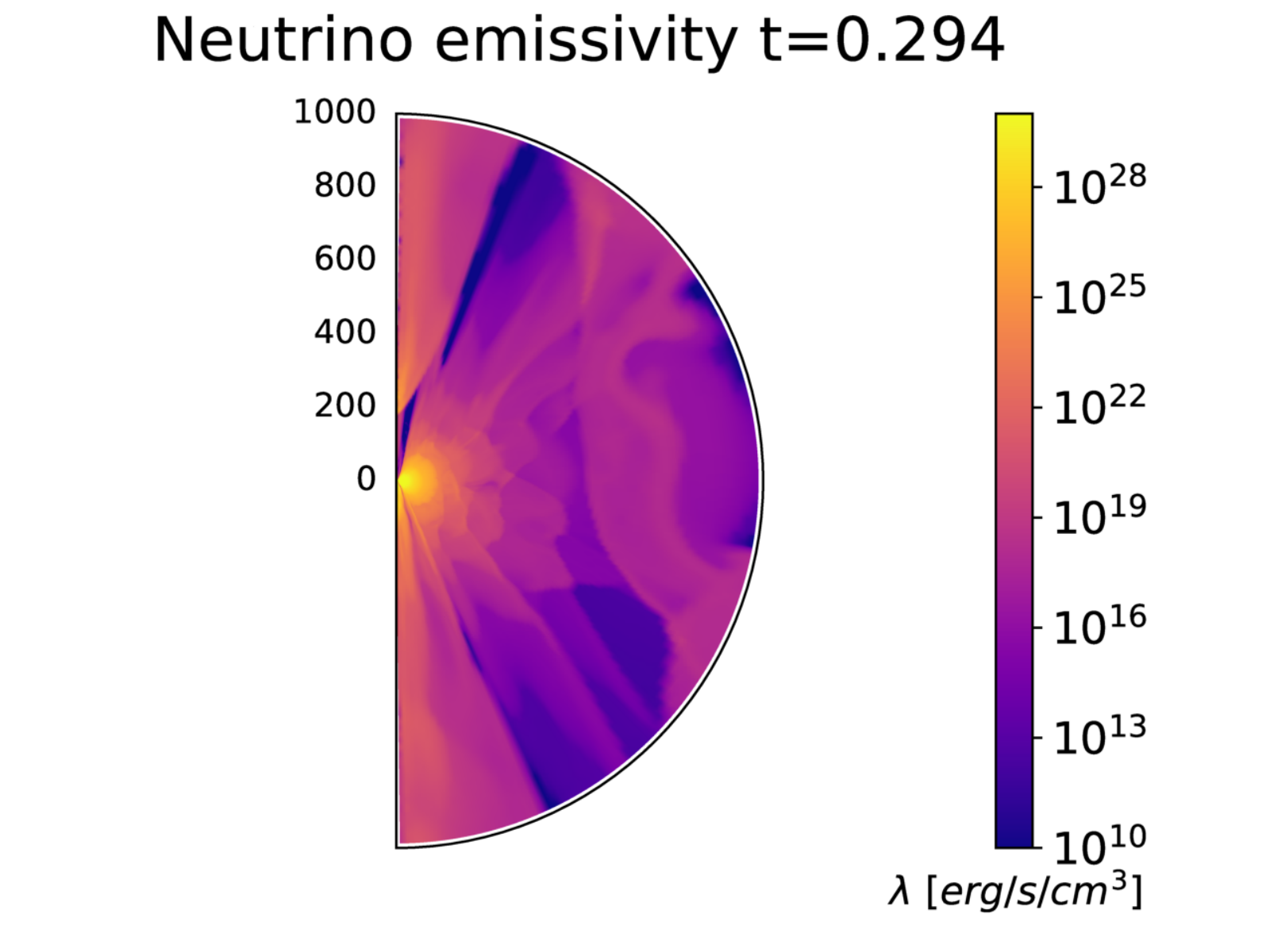} &
  \hspace{-12mm}\includegraphics[width=0.39\textwidth]{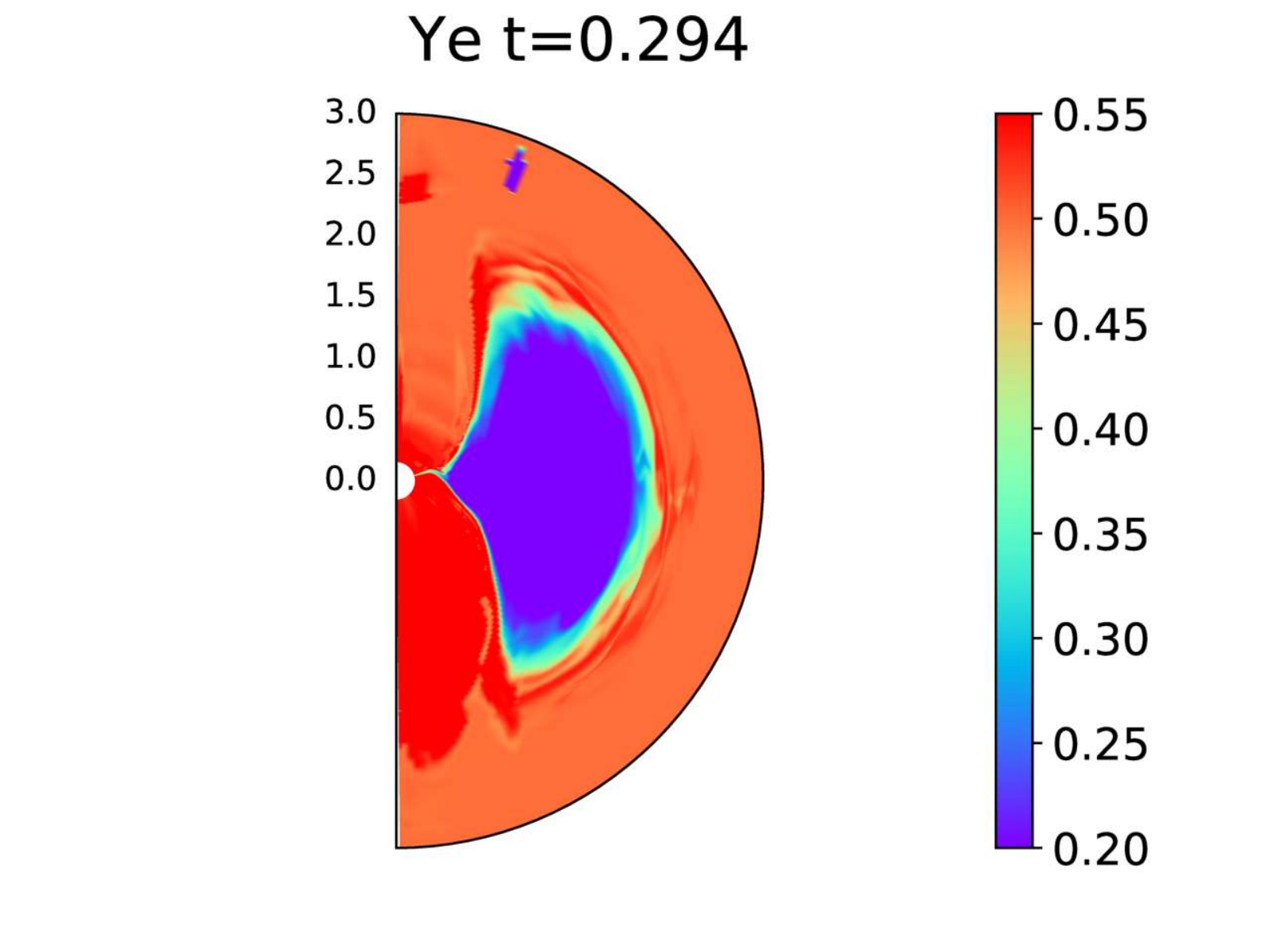}
\end{tabular}
\caption{The results, from left to right, for the \textit{Density and Magnetic field, Neutrino emissivity, Electron fraction} distributions, for four models. 
The color maps are taken at the end of the simulation, evolved until time $t_{f}=20000~M$ (i.e., $\sim 0.3$ second for $M_{\rm BH}=3 M_{\odot}$). The models from top to bottom, are
LS-Therm, LS-Magn, HS-Therm, HS-Magn. Note, that in the first two columns we use the linear scale in radius, while in the last column the logarithmic scale is used.
}
    \label{fig:models_results}
\end{figure*}

In Figure \ref{fig:models_results} we show the profiles of density, magnetic field, neutrino emissivity, and electron fraction, in the $r-\theta$ plane, as obtained at the end of each simulation. The models, from top to bottom row, represent the cases of LS-Therm (low spin a=0.6, large gas to magnetic pressure ratio, $\beta=100$),
LS-Magn (low spin a=0.6, smaller gas to magnetic pressure ratio, $\beta=10$), HS-Therm (high spin a=0.9, large gas to magnetic pressure ratio, $\beta=100$), and HS-Magn (high spin a=0.9, smaller gas to magnetic pressure ratio, $\beta=10$). As can be noticed, the more magnetized models have an effect on the larger extension of the magnetically driven and neutrino-cooled winds
at the equatorial plane and intermediate latitudes. These winds are moderately dense ($\rho\sim 10^{5}-5 \times 10^{8}$ g cm$^{-3}$) and hot ($T \sim 10^{9}-3\times 10^{10}$ K).
In addition, there are hot luminous regions near the black hole rotation axis, which have a very low baryon density. Their neutrino emissivity is correlated with the black hole spin value (see e.g. also \citet{2016PhRvD..93l3015C} for the neutrino emissivity of the disk dependent on the BH spin).

The $Y_{\rm e}$ value is determined in the grid at every time step (from Equation \ref{eq:yee}).
The electron fraction is very low ($Y_{\rm e}\sim 0.1-0.2$) only in the very central, innermost parts of the accretion torus. The wind material follows
the trajectories
that have this initial value of $Y_{\rm e}$, and reach the outer boundary during the dynamical simulation.
Notice that in the electron fraction maps shown in Figure \ref{fig:models_results}
we use the logarithmic scale in radius, as most of the neutronised matter is located below 100 $r_{\rm g}$. In the outflows that reach outer boundary, the thin filaments
(visible in the maps of magnetic field distribution, made in the linear scale)
drive outwards the neutronised matter. However, the  bulk of background material (with very low density) has the final electron fraction of $Y_{\rm e} \sim 0.5$.

\begin{figure}
    \includegraphics[width=0.5\textwidth]{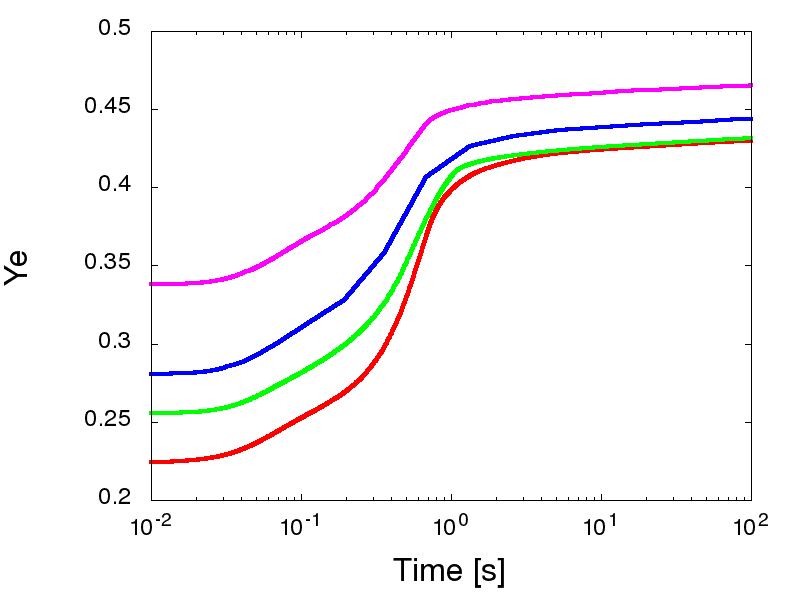}
    \caption{Time dependence of the angle-averaged
      electron fraction in the outflowing material. Different colors present models HS-Therm (magenta), HS-Magn (blue), LS-Therm (green), and LS-Magn (red) }
    \label{fig:yetime}
    \end{figure}

  The electron fraction in the outflow rises with time (see Figure \ref{fig:models_results}).
  However, some earlier outflows must have much lower Ye in order to produce the 2nd and 3rd peak elements.
  The time dependence of the electron fraction is presented in Figure \ref{fig:yetime}, as measured along the tracers and averaged over all angles.
  As shown in the plot, during the first second the average
  electron fraction of the ejecta is between 0.2, and 0.35. It rises then up to above $Y_{\rm e}=0.45$, as the outflows expand.
  The trends observed for the four models considered are such that the more magnetized outflows are on average more neutron rich, for the same black hole spin.
  Also, this angle-averaged distribution
  implies that the low spinning black hole produce outflows that
  are on average more neutron rich.
 
The temperature maps are not shown, but the distribution of temperature is directly
followed by the neutrino emissivity (the cooling rates for the neutrino
emission by URCA, electron-positron anihillation, bremsstrahlung, and plasmon process, scale with temperature in the power 6, 9, 5.5, and 9, respectively, see
\citet{Janiuk2004MNRAS}).

The wind density can be visualized also by means of the outflow tracers distribution, as was shown in Figure \ref{fig:in_beta}.
We notice that despite of the fact that all these tracers were uniformly distributed in the initial state, their final distribution strongly
depends on the model parameters.
For a higher BH spin value, much fewer tracers particles are detected
in the
wind outflows, than for a low BH spin, if the flow is more magnetized. In contrast, for
the thermally-dominated models, the higher BH spin helps launching denser wind outflows. This fact can be understood in terms of competitive action of the MHD turbulence
for the winds acceleration, and the Blandford-Znajek process. The latter is
strongly dependent on the BH spin value so that for almost maximally rotating black holes the B-Z process will overcome on the uncollimated outflows (see \citet{Sap2019ApJ}).

\begin{figure}
    \centering
     \includegraphics[width=0.4\textwidth]{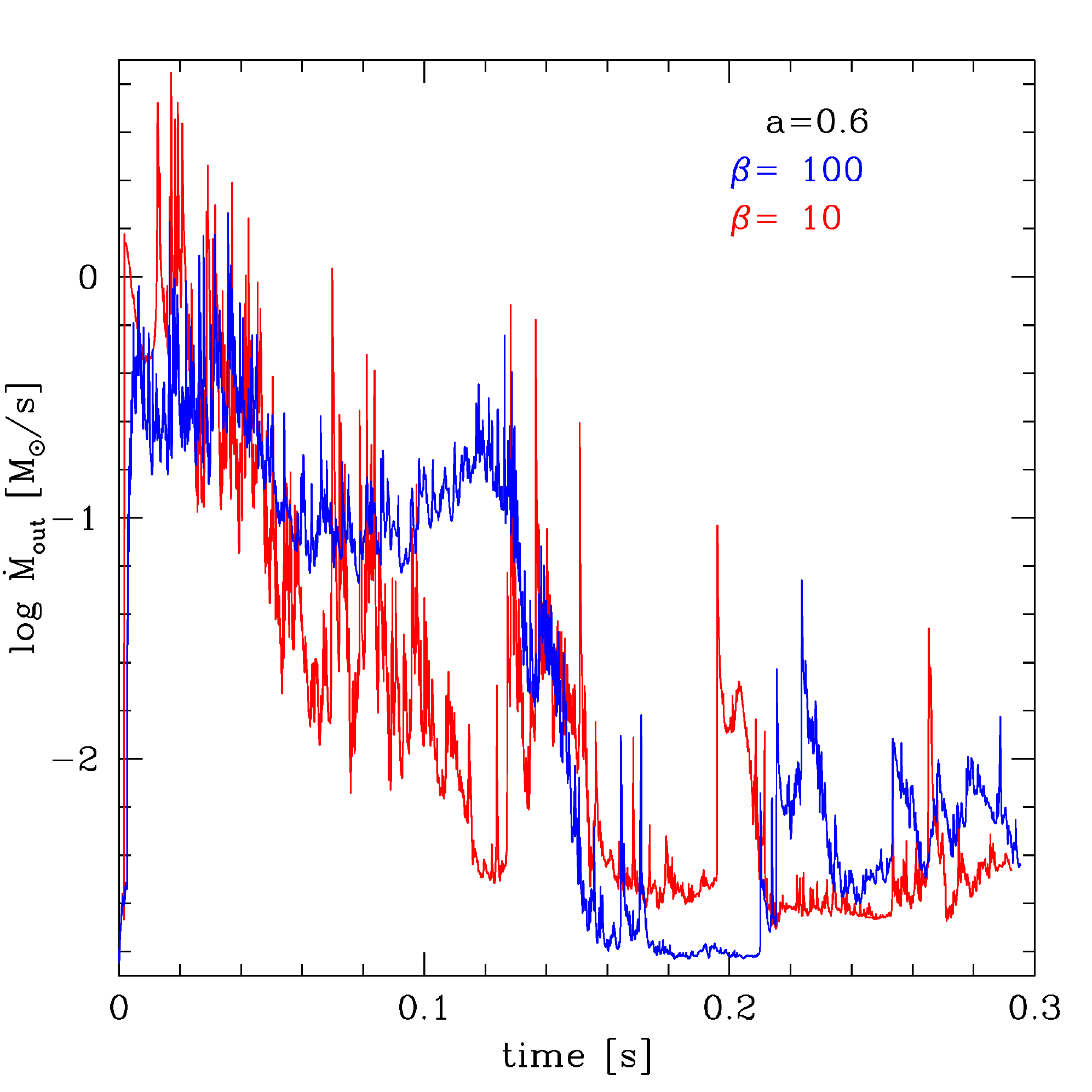}
     \includegraphics[width=0.4\textwidth]{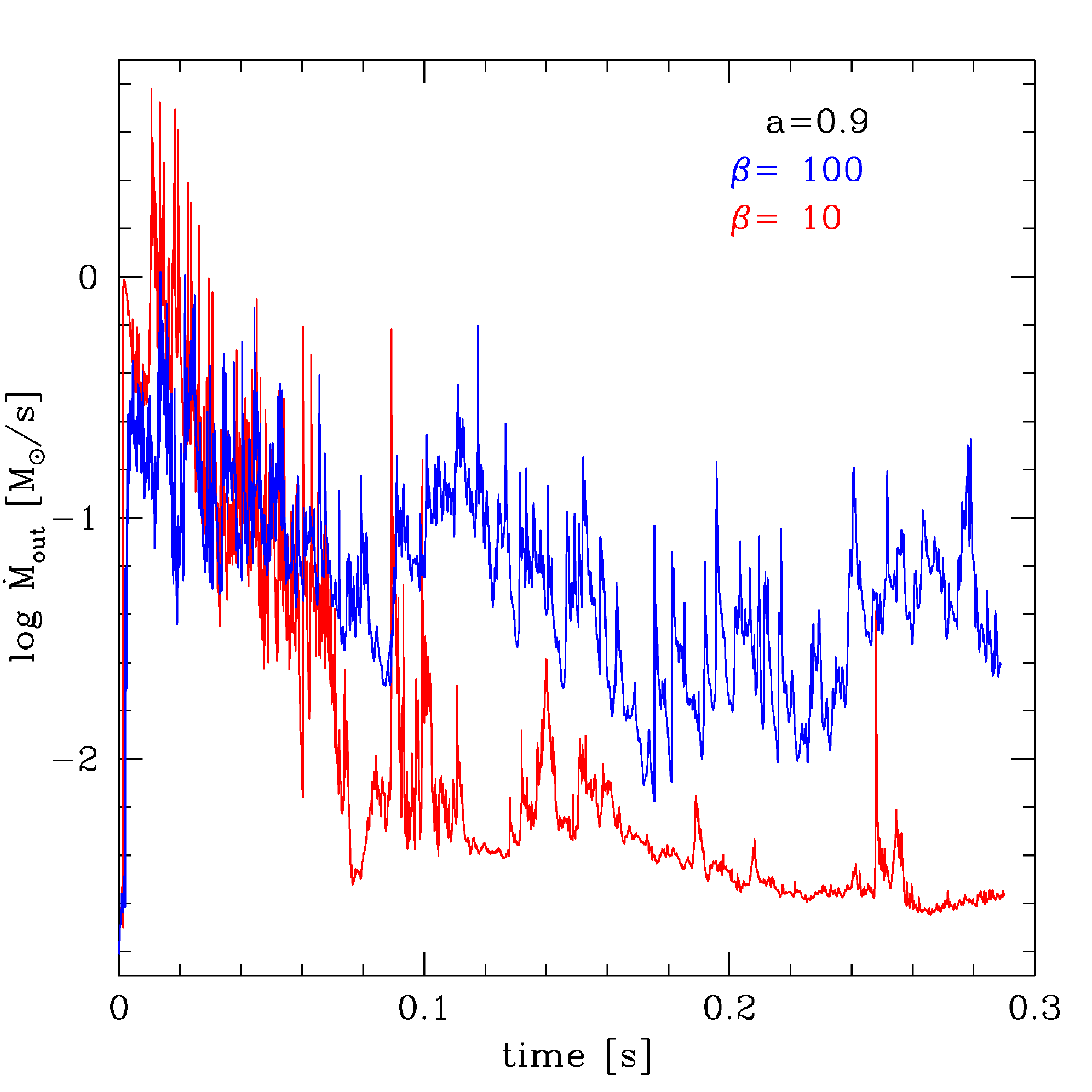}
     \caption{Time dependence of the mass flux through the outer boundary of the torus simulation. Blue lines denote the models with magnetisation $\beta=100$ while the red lines denote $\beta=10$. Two black hole spin values are a=0.6, and a=0.9, as shown in the left and right panels, respectively}
     \label{fig:mdotout}
     \end{figure}

In Figure \ref{fig:mdotout} we show the time dependence of the mass
loss rate
 through the outer boundary in the function of time during our simulations.
 As shown in the figure, the mass outflow is huge at the beginning of the simulation, when the initial condition of the pressure equilibrium torus is being relaxed. However, after some 0.03 seconds (which corresponds to about 2000 M), the outflow rate saturates at a small value. At the end of the simulation, the outflow rate is below 0.05 $M_{\odot}$s$^{-1}$.
 The models with more magnetic pressure result in higher time-averaged mass loss rates (see Table \ref{tab:in}).
   Also, the total mass lost from the outer boundary follows the same trend with magnetisation.
 For large BH spin of a=0.9,
 the mass outflow rate in the second half of the simulation is larger
  in less magnetized model, while initially it was smaller.
  The effect seen in the Figure
  might be partially an artifact of the initial condition, and
  the MRI turbulence decay at late times.
  We notice, that
  the total mass that is lost from from the disk during our simulations
  is between 17\% and 38\% of the initial disk mass (see Tab. \ref{tab:in}).
  This number however takes into account both the unbound outflows, and mass
  accreted through the BH horizon,
  and also the polar jets (albeit these are of a very low density).
  The mass lost through the outer boundary appears to be in the range
  of 2\%-16\% only, in contrast to the results of \citep{Fernandez2019}.
  The instantaneous mass of the unbound outflowing material, as estimated via sampling the tracers, is also consistent with the above results and with the intuitive prediction that larger magnetic field helps driving larger mass of the outflows (see Table \ref{tab:therm}).
  Finally, we checked for the amount of unbound matter with $-h u_{t}>1$, i.e.
  the condition corresponding to a positive Bernoulli parameter
  in Newtonian gravity.
  It results in the potential mass of the outflows
 in the range between $10^{-3}-10^{-2} M_{\odot}$, varying with time in the simulation,  with higher values obtained for more magnetized models.

\subsection{Nuclear reaction network}

We use the data from our simulations, and the particle trajectories, as an input to the nuclear reaction network. These computations are performed using the code SkyNet \citep{2017ApJS..233...18L}, in the post-processing simulation. The code is capable to trace nucleosynthesis in the rapid neutron capture process and involves large database of over a thousand isotopes. It takes into account the fission reactions and electron screening.
In our SkyNet runs, the Helmholtz EOS is used by this code, and it calls for the weak and strong reaction libraries, and for the spontaneous and symmetric fission. Self heating is taken into account, while the screening is neglected.

The results of MHD simulations that are stored in the tracer particles data are
the density, temperature and electron fraction. 
In the post-processing, the density distribution is followed along each of the particle trajectories,
and the piecewise linear function is used for interpolation of the densities.
When the time is exceeding the original simulation timescale,
the density profile is extrapolated with a power-law continuation of $t^{-3.0}$, in a good agreement with the homologous expansion of the outflow.
Because the nuclear self-heating is taken into account,
we take only the initial value of temperature on each trajectory,
and then the temperature adjusts itself to the reactions balance.
We checked that the average
temperature in the outflows rises during the first $\sim 10$ seconds,
and then drops to below $10^{6} K$ after several hundred seconds
from the outflow launching.
The electron fraction value is also read from each of the trajectories initially, to establish the conditions for nuclear statistical equilibrium and
initial chemical abundance pattern.
When following the outflow, the electron fraction is
updated according to the nuclear reactions balance.
The values less than $Y_{e}\sim 0.05-0.3$ are found in the tracers
below 1 second of expansion. After hundred seconds, the electron fraction
level in the wind saturates, and stays around $Y_{e}\sim 0.4-0.5$,
depending on the angle.

\begin{figure}
    \centering
    \includegraphics[width=0.3\textwidth]{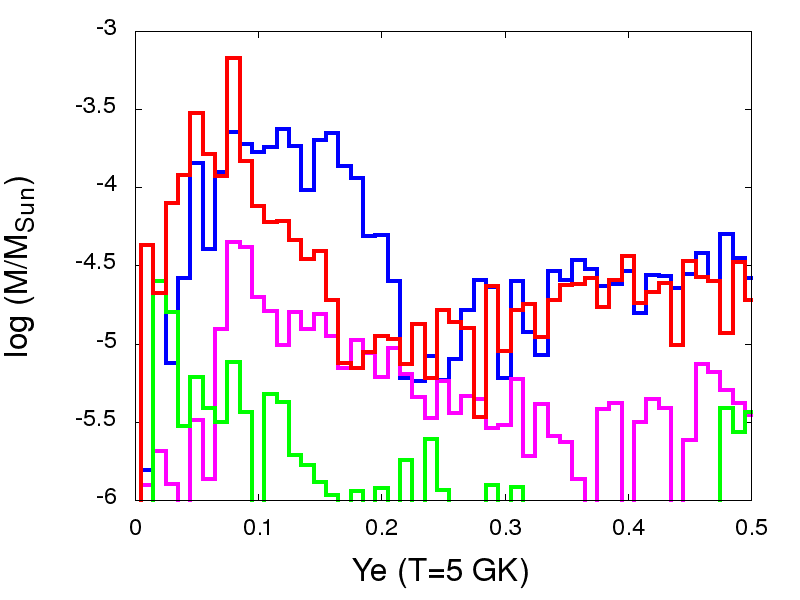}
    \includegraphics[width=0.3\textwidth]{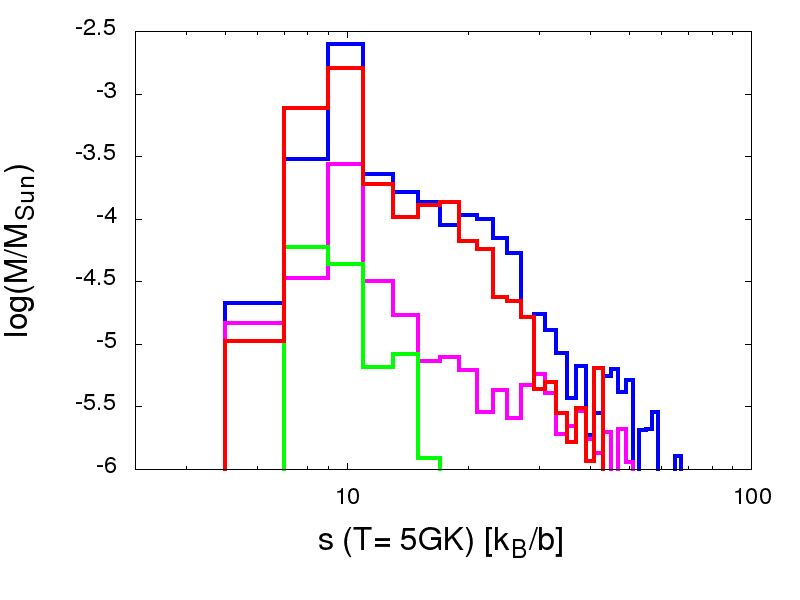}
    \includegraphics[width=0.3\textwidth]{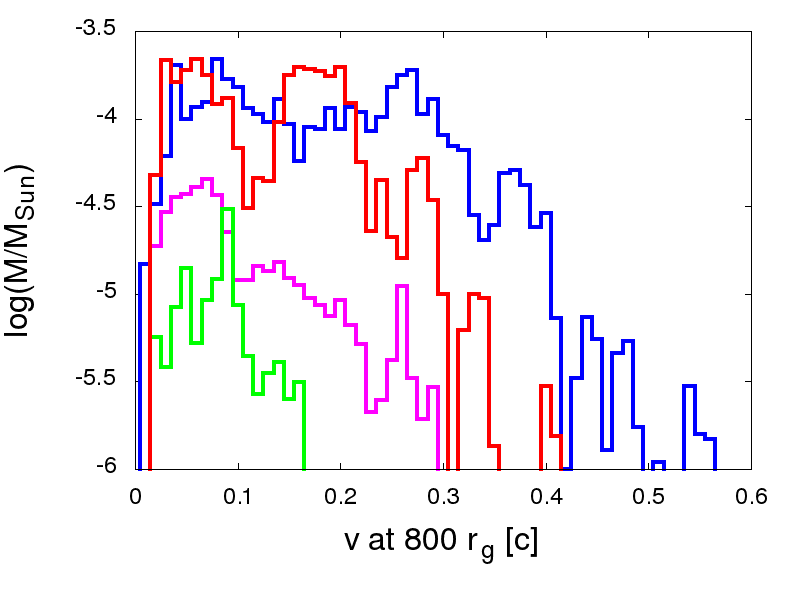}
    \caption{Mass distributions of the unbound disk outflows of $Y_{\rm e}$, entropy, and velocity. Plots show the electron fraction and entropy as measured at the outflow trajectories in the region where the temperature drops to 5 GK, and the velocity at trajectories measured in the distance of 800 $r_{g}$ (i.e. $\sim 3560$ km). With different colors we present models HS-Therm (magenta), HS-Magn (blue), LS-Therm (green), and LS-Magn (red)   }
         \label{fig:histograms}
    \end{figure}

  To probe the nucleosynthesis process in our simulations, we investigated the thermodynamic properties of the outflowing ejecta.
  In Figure \ref{fig:histograms} we present histograms of the electron fraction, entropy, and velocity, as distributed according to the mass carried in the outflows, showing how much mass within the outflows carries these quantities of certain values. The distributions are plotted at the time, when the outflow temperature is still large, and equal to $5 ~GK$.
  As shown in the first panel of this Figure, the largest mass of the outflow with smallest electron fraction, $Y_{\rm e}<0.2$, is launched in the simulations with stronger magnetic field. Here, the comparison between HS-Magn, and LS-Magn histograms, reveals the additional influence of the black hole spin on the results. The $Y_{\rm e}$ distribution of highly neutronized outflows
  is narrower for the spin $a=0.6$, than for $a=0.9$. It means that fastly rotating black holes tend to launch slightly less amount of the most neutron rich outflows, while the overall mass of the outflow with $Y_{\rm e}>0.2$ remains similar. These outflows have also broader distribution of entropy, and velocity (middle and right panels).
  
  For simulations with the weak magnetisation, both the total mass of the outflow, and the fraction of mass with $Y_{\rm e}<0.2$ is very small, and it is
  smaller for the lower black hole spin.
The specific entropy in these outflows is concentrated around 10 $k_{B}/b$, and their velocity does not exceed $0.3~c$ at the distance $800 ~r_{\rm g}$.
The mean values of the outflow velocity, electron fraction, and entropy, are summarized in Table \ref{tab:therm}.

\begin{figure}
    \centering
    \includegraphics[width=0.4\textwidth]{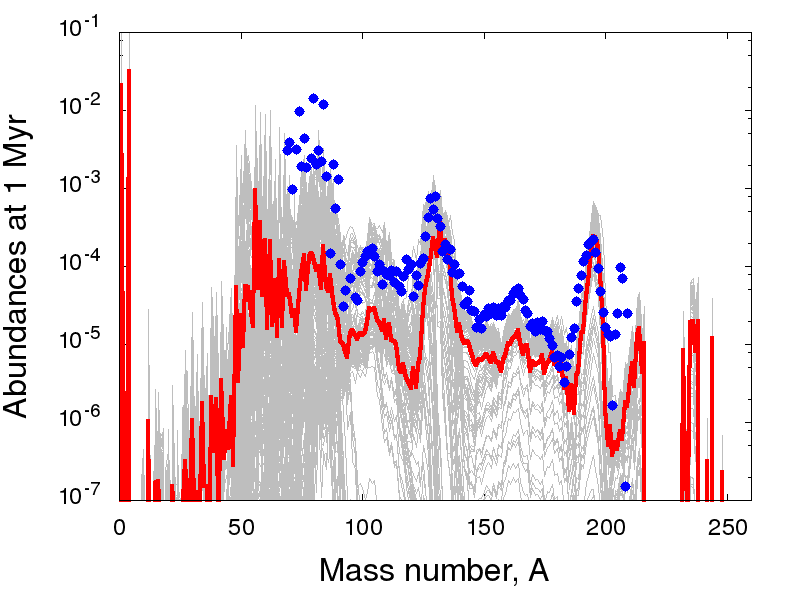}
    \includegraphics[width=0.4\textwidth]{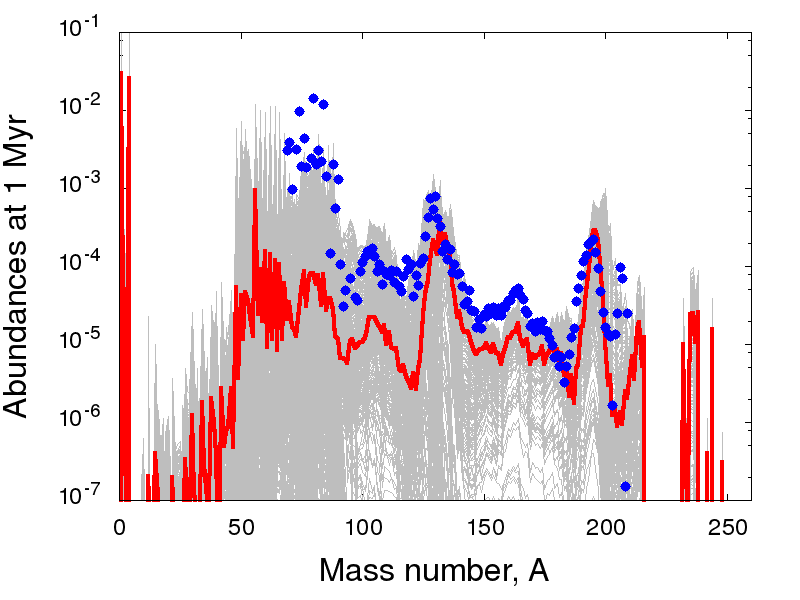}
    \includegraphics[width=0.4\textwidth]{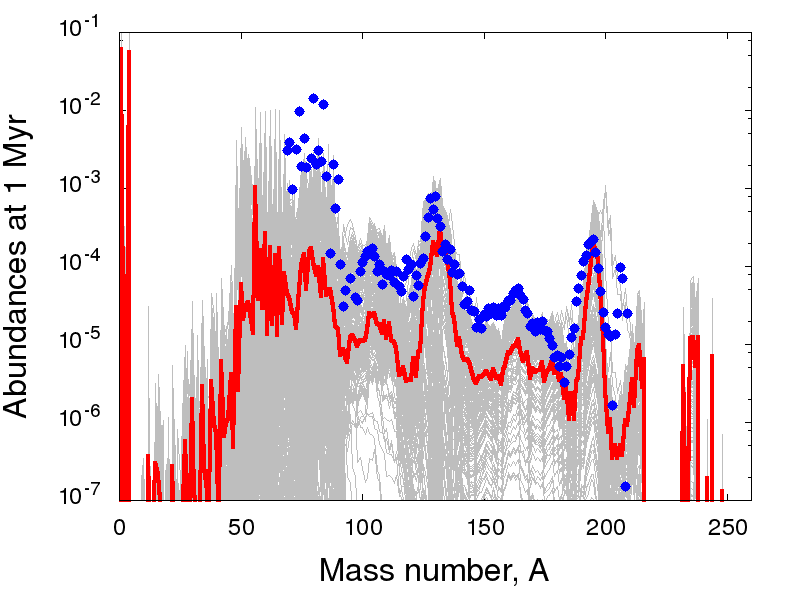}
    \includegraphics[width=0.4\textwidth]{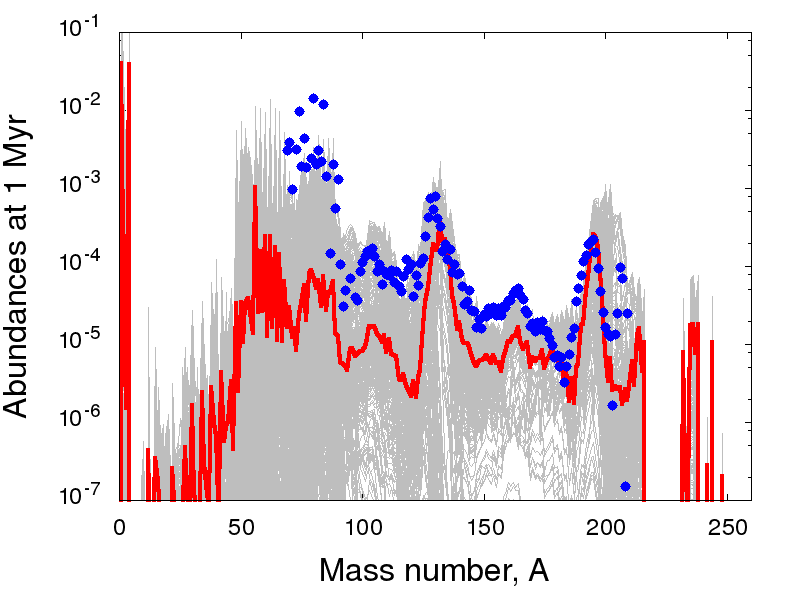}
     \caption{The results of element relative abundance as a function of the mass number, calculated for tracer particles on the outflow from the accretion disk simulation, and extrapolated until time $t=1 Myr$. Parameters of the models are black hole spin $a=0.6$ (top) and $a=0.9$ (bottom), and plasma magnetisation, $\beta=100$ (left) or $\beta=10$ (right)} 
     \label{fig:in_phase}
\end{figure}

In Figure \ref{fig:in_phase} we show the final results of r-process nucleosynthesis in our simulated black hole accretion disk outflows.
Shown are
four models, for two values of the black hole spin $a=0.6$, and two values of the gas-to-magnetic pressure ratio, $\beta=100$ (left panel), and $\beta=10$ (right panel). The results for the relative abundances of created isotopes are obtained here after
extrapolation of the wind outflow up to the time of 1 Myr, and compared
with the Solar abundance data taken from \citet{2007PhR...450...97A}.

We note that the second and third peaks of the r-process elements distribution
are well reproduced in our simulated profiles.
The relative abundances in the first peak are however somewhat smaller
than observed. This is found
at most trajectories (grey lines), and in the profiles averaged
over all trajectories (red line). This
would suggest that the Iron group isotopes can be
synthesized in the short GRB accretion disks outflows, but only at
moderate rates.

In the network simulations, we can trace the angular distribution of the outflow
trajectories, on which
the elements of subsequent abundance peaks are produced.
We found that the most efficient production of the second and third peak elements,
$A\sim130$ and $A\sim200$, is concentrated
on the polar angles $70^{\circ}<\theta<120^{\circ}$, so roughly around the equator.
For the first peak, the most abundant production of the elements with $A\sim60$ is found
at trajectories below $\theta\sim 70^{\circ}$ and above $\theta\sim 120^{\circ}$, so close to the polar axes.
Because these trajectories give only a moderate contribution to the total mass
outflow, the
averaged abundance pattern is under-representing the first peak magnitude,
relatively to the second and third one, as seen in Fig. \ref{fig:in_phase}.
The geometrical dependence of the abundance pattern could potentially reveal a
higher magnitude of the first peak, in some specific cases of GRB-related kilonovae as  observed closer to the polar axis.

As for the relative distributions of the elements with mass number
$A\sim 130$ (Xenon group), their profiles resulting from our simulations match the Solar data more closely for the short GRB models with
higher-magnetized winds,
and a less spinning black hole in their engine.
This might tentatively suggest that such black holes should be more frequently produced
via the binary neutron star mergers,
which is in agreement with the moderate spins of  black
holes resulting from the simulation
of hypermassive neutron star that undergoes delayed collapse
\citep{Ruiz2016ApJ}.
We intend to investigate further the problem of the black hole spin
influence on the kilonova signal in the future work, implementing the 3-dimensional
scheme for the MHD turbulence to disentangle
the effects of magnetic driving and BH rotation.

\begin{deluxetable}{lcccc}

\tablecaption{Thermodynamic properties of the outflows \label{tab:therm}}

\tablehead{
  \colhead{Model} 
  & \multicolumn{1}{p{3cm}}{\centering  Unbound outflow mass [$M_{\odot}$] }
  & \multicolumn{1}{p{2cm}}{\centering average $Y_{e}$ \\ at $T=5 GK$} 
  & \multicolumn{1}{p{3cm}}{\centering average $s$ [$k_{B}/b$] \\ at $T=5 GK$ } 
  & \multicolumn{1}{p{2cm}}{\centering average $v$ [$c$] \\ at $800 r_{\rm g}$ } 
}
\startdata
      HS-Therm & 0.00043  & 0.34 & 17.57 & 0.151  \\
      HS-Magn & 0.00386   & 0.28  & 14.56 & 0.229 \\
      LS-Therm & 0.00012  & 0.26 & 12.83 & 0.110  \\
      LS-Magn & 0.00315    & 0.22  & 12.89 & 0.177 \\
\enddata
\tablecomments{Model parameters are given in Table 1. The total mass, angle-averaged electron fraction, and entropy of the outflows are computed for the tracers in the unbound material that has cooled down to temperature of 5 GK. The velocity in the units of speed of light is measured at distance of 800 gravitational radii from the black hole}
\end{deluxetable}

\section{Summary and conclusions}

We modeled the nucleosynthesis in black hole accretion disks and outflows
at the central engine of a short gamma ray burst. The result is abundant production of light isotopes, with mass numbers in the range $A\sim 60- 80$, which corresponds to the first maximum of nuclide production in the r-process.
These nuclei have been found in the simulations of the
prompt phase of the GRB under the statistical equilibrium conditions
\citep{Janiuk2017}.
In the present simulations we found 
that also heavier isotopes are produced, up to the mass number $A\sim 200$, and are created on the outflow trajectories that start from the surface of an accretion disk and in the area beyond it. These outflows are carried
by the magnetized, neutrino-cooled wind.

The magnetic fields are mainly responsible for the transport
of the angular momentum, which enables accretion, but is also 
driving the disk outflows.
The specific choice of the magnetic field configuration is not new,
as we adopt the simple poloidal configuration with the field lines which follow the iso-contours of constant density. This is commonly adopted,
because a crucial component of any electromagnetic jet launching model is the presence of a poloidal field component \citep{Beckwith_2008, Paschalidis_Ruiz_Shapiro_2015}. 
Such configuration allows the evolution of the BH magnetosphere and
formation of the large scale open field lines, along which the Blandford-Znajek process may extract the rotational energy of the BH. Only to some extent, the neutrino annihilation can be a complementary process to power the GRB jets \citep{Mochkovitch_et_al_1993, Aloy_et_al_2005, Janiuk2013, Liu2015, Janiuk2017}.

In our simulations, the number of outflowing trajectories and the average mass outflow rate in case of higher 
magnetic pressure is much larger than for small magnetisation (for the same black hole spin).
We verified with a test simulation (assuming the adiabatic EOS), that
the models with no magnetic fields produce essentially no outflows.
Clearly, such models should preserve the stationary solution
of the FM torus in equilibrium. In fact, after time $t_{f}=20000 ~M$, the number of the outflow trajectories in the model with no B fields,
$\beta_{0}=\infty$, and 
BH spin of $a=0.6$, was only
$N_{\rm out}=47$. We treat this result as a numerical artifact.
(We checked that
 models with no magnetic field, but equipped with numerical EOS and neutrino
 cooling functions, do not produce outflow trajectories 
that leave the
 computational domain at $R_{\rm out}\sim 800-1000 r_{g}$.)
For comparison, the adiabatic model but endowed with initial poloidal field with large plasma beta parameter $\beta_{0}=50$, produced almost 20 times larger number of outflow tracer particles, $N_{\rm out}=724$, and the one with $\beta_{0}=10$ 
resulted in $N_{\rm out}= 1683 $ trajectories in outflow.
However, many more trajectories and $N_{\rm out}=2517$ were found in the
fiducial model LS-Magn.
This confirms, that not only the magnetic fields, but also the microphysics in our simulations
is responsible for driving the outflows.
In the present computation, because only the neutrino cooling is
accounted for, and neutrino heating is missed, we attribute this
effect rather to the chemical composition of the torus. The Helium nuclei, and their photodissociacion, may be a source of extra
energy generation and hence also the outflow drivers.

The wind material is contributing to the nucleosynthesis in the
outflowing ejecta, and the outcome of this process depends on the magnetic field strength. The full 3-dimensional simulation is a next step that will 
 verify the nucleosynthesis yields dependence
 on the field configuration and the  BH parameters.
 Still, even in this axisymmetric simulation of the GRB engine, we have some residual magnetic turbulence, which drives the outflow.
 We show that the results
 of current simulation
 somewhat depend on the value of black hole spin.

 Our study  is focused on the variation of BH spin and initial magnetization, and in addition to the recent works devoted to the post-merger disks
  \citep{Fernandez2015, Fernandez2019, miller19} we indicate that it is
  the high magnetisation of the torus, with a help of rapidly spinning black hole, which is likely to produce
  a wide range of neutronisation in the outflows (the $Y_{\rm e}$ in the range of 0.1-0.45).
  Only for the low-spin, thermal model LS-Therm, our obtained 
  electron fraction values in the unbound ejecta are around $Y_{\rm e}<0.2$,
  which leads to a 'red kilonova'. Otherwise, the spin of the black hole $a=0.9$, with the same magnetisation, results in a contribution of higher $Y_{\rm e}$ ejecta with 0.2-0.4.
  The crucial role of magnetic field in producing the broad range of $Y_{\rm e}$
  is in line with the findings of \citet{Fernandez2019}, who conducted both the GRMHD models and the pseudo-Newtonian $\alpha$-disk simulations. The results are somewhat quantitatively different, possibly because of differences in numerical scheme regarding the implementation of the equation of state.
 
  Our wind ejecta have higher average velocities, which for the HS-Magn model
  reaches $0.23 ~c$,  while the mass in these ejecta is rather small.
  The estimated mass of the outflow in the unbound
  tracers is in agreement with the mass of the flow with
  a positive Bernoulli parameter. The cumulative mass lost via the outflow
  over the simulation time can reach even 17\% of the torus mass, and contain the lanthanide-poor component,
  if the magnetisation and spin of the black hole are high.  
  It remains to be checked whether a more elaborate neutrino transport scheme, and different treatment of scattering,
  would change significantly
  our results. We can speculate now that the neutrinos
  should help launching a more 'blue kilonova', for a smaller black hole spin than probed in this work, as discussed e.g. in \citep{miller19}.

The open question remains still about the observational verification of present simulations. Do these outflows provide
enough mass to be detectable in the kilonova lightcurves via the fits to
their continuum emission? Can the specific emission lines be detected, and help verify if the
radioactive material comes from the dynamically disrupted expanding
tails that formed prior to the NSNS merger, or from the accretion disk winds?
The previous works on the gamma-ray bursts accretion disks have shown already
 \citep{2006ApJ...643.1057S}, that the large overproduction factors 
in their outflows occurs for $^{44}Ti$, $^{45}Sc$, and $^{64}Zn$. 
Also, \citet{2003NuPhA.718..611F} found that the light p-nuclei, such as $^{92}Mo$, and $^{96}Ru$, and $^{114}Sn$  are produced in the rapidly accreting disks.
The neutrino-driven winds modeled by \citet{2014MNRAS.443.3134P} for the short GRB progenitors, have shown that the wind can contribute to the weak r-process in the range of atomic masses from 70 to 110.

In the present work, we show that the MHD driven winds from the accretion disks in GRB engines
can contribute to the further peaks of the r-process nucleosynthesis,
beyond $A\sim 130$.
Potentially, the emission from the radioactive decay of these species can
give a separate, observable signal in the kilonova lightcurves. Its modeling requires however full radiation transport computations.
Since the accretion disk wind is rather light, but can reach the velocities
higher than about 0.1-0.3 $c$, it can catch up with the precedent dynamical
ejecta and form a separate component.
Its contribution would be
visible in the bluer part of the optical g-band,
and is currently not accounted for in the modeled
lightcurves, as presented e.g. in \citet{2017ApJ...848L..17C}.

As for the overall chemical enrichment of the inter-stellar medium due to the
bulk action of short GRB populations, there have been many claims for the dominant role of neutron star mergers. For instance, for the heavy elements, such as Gold and Europium, it has been shown
recently that NS-NS mergers can produce it in sufficient amounts and are likely to be the main r-process sites \citep{2018ApJ...855...99}.
However, the nucleosynthesis predictions of the current core-collapse supernova models are in agreement with Fe-group yields and
trends in alpha element to Iron ratio \citep{2019ApJ...870....2C}.
Form our simulations, we also conclude that bulk of the elements with $A\le 70$ in the Solar system,
should rather be of a different origin than the short GRB central engine outflows,
however the dynamical ejecta from the disrupted neutron star prior to the GRB event cannot be excluded in this context.

\section*{Acknowledgments}
We thank Jonas Lippuner, and Oleg Korobkin, for helpful discussions.
 We also thank the anonymous referee for detailed comments and suggestions that helped us to improve our manuscript.
This work was supported in part by the grant 
no. DEC-2016/23/B/ST9/03114, from the Polish National Science Center.
We also acknowledge support from the Interdisciplinary Center for Mathematical Modeling of the Warsaw University, through the computational grant Gb70-4, as
well as the PL-Grid infrastructure under project grb-2.

\bibliographystyle{aasjournal}
\bibliography{paper_gold_v2.bib}

\end{document}